\documentclass[
preprint,
 amsmath,amssymb,
 aps,
]{revtex4-1}

\usepackage{graphicx}
\usepackage{dcolumn}
\usepackage{bm}
\usepackage{txfonts}
\usepackage[T1]{fontenc}

\def\i{\mbox{\rm i}}
\def\be{\begin{eqnarray}}
\def\ee{\end{eqnarray}}

  \makeatletter
    
    \@addtoreset{equation}{section}
  \makeatother

\begin{document}


\title{Drag Coefficient of a Circular Inclusion\\
in a Near-Critical Binary Fluid Membrane}

\author{Hisasi Tani$^{1,2}$}
\email{Email: htani0926@tamu.edu} 
\author{Youhei Fujitani$^2$}
\email{Email: youhei@appi.keio.ac.jp}
\affiliation{$^1$ Department of Mechanical Engineering, Texas A\&M University, TX 77843-3123,
USA \\ $^2$ 
School of Fundamental Science and Technology,
Keio University, 
Yokohama 223-8522, Japan} 
\begin{abstract}
We calculate the drag coefficient of a circular liquid domain, 
which is put in a flat fluid membrane composed of a binary fluid mixture 
lying in the homogeneous phase near the demixing critical point. 
Assuming a sufficiently small correlation length, we regard the domain dynamics
as independent of the critical fluctuation and use the Gaussian free-energy functional
for the mixture.  Because of the near-criticality, the preferential attraction between 
the domain component and one of the mixture components generates 
the composition gradient outside the domain significantly and can affect the drag coefficient.
We first consider a domain having the same membrane viscosity as the domain exterior.  
The drag coefficient is expanded with respect to a dimensionless strength of the preferential attraction.  
It is numerically shown that
the magnitude of the expansion coefficient decreases much as 
the order of the strength increases and that the first-order term of the series usually gives a good approximation 
for practical material constants.  The effect
of the preferential attraction is shown to be able to become significantly large in practice.  
We second consider cases where the membrane viscosities of the domain interior and exterior are different. 
The first-order term of the expansion series decreases to approach zero
as the domain viscosity increases to infinity.  
This agrees with previous numerical results
showing that the hydrodynamics makes the effect of the preferential attraction negligibly small for a rigid disk. 
\end{abstract}

\maketitle
\section{\label{sec:intro}Introduction}
A colloidal particle moving 
translationally with a sufficiently small speed
in a quiescent fluid suffers a drag force whose
magnitude is proportional to the speed.
The constant of the proportionality is called the drag coefficient, and 
can be related to the self-diffusion coefficient \citep{suther, eins, 111years}.  
The Brownian motion of a particle gives some informations on the properties of the medium.
This kind of probing experiments have been done widely in the microrheology \citep{MicroRev, viscel2}, and some 
have been done for the fluid membrane \citep{glaz,ortega}. The Brownian motion of a trapped particle is now
detected with high resolutions of approximately $1\ $nm and $1\ \mu$s \citep{nat,grim}.   \\

The drag coefficient of a circular inclusion in a fluid membrane has been studied extensively. 
A typical example of a fluid membrane is 
a lipid-bilayer membrane contained in the biomembrane \citep{singer, andel}.  Regarding
a membrane protein as a rigid disk in a flat two-dimensional (2D)  fluid immersed in
a three-dimensional (3D) fluid,  Saffman \& Delbr{\" u}ck calculated its drag coefficient by applying
the Stokes approximation \citep{saffman}.  The calculation was later performed more thoroughly \citep{saffman2,hughes}. 
The raft hypothesis, which asserts that microdomains enriched in specific lipids,
should give platforms to biochemical reactions \citep{ikonen, leslie}, triggered experimental studies on the phase separation
of artificial multicomponent membrane \citep{Dietrich, Veatch2002, VeatchBJ, keller, veatch1}.  
The dispersed phase can take a distinct 
circular shape, {\it i.e.\/}, a circular liquid domain is realized in a fluid membrane.  In \citet{koker},
the drag coefficient of a domain in a flat fluid membrane
was calculated on the assumption that the membrane viscosities are the same in the domain interior and exterior.
Some researchers measured the diffusion coefficient of a circular liquid domain,
and analyzed the results by using the theoretical result for a rigid disk \citep{cicuta} or by assuming 
that the membrane viscosities are the same in the domain interior and exterior \citep{alias}. 
The drag coefficient of a circular liquid domain with a distinct membrane viscosity
in a flat fluid membrane has recently been calculated \citep{confine, tani}.
Multicomponent membranes near the demixing critical point were also studied experimentally.
The observed static critical exponents were found to agree
with the ones of the 2D Ising model \citep{honerBJ, veatch2, honer}, while the observed dynamic
critical exponent turned out to be explained in the framework of the model H  -- a standard model
for the near-critical dynamics \citep{Hohenberg, Onukibook} -- with the dynamics in the ambient 3D fluid
being taken into account \citep{inafuji, haat, prl, hydro}.  \\

Suppose that a circular inclusion is 
put in a flat fluid membrane lying in the homogeneous phase
near the demixing critical point \citep{yeth, demery, camley}. 
Components of the mixture are usually attracted unequally by the inclusion, and thus
the composition gradient is generated significantly 
around the inclusion in the near-critical fluid membrane. 
An inclusion moving translationally suffers the drag force
exerted by the ambient 2D and 3D fluids.  Generating the osmotic pressure in the membrane,
the composition gradient can alter the flow fields and affect the self-diffusion of the inclusion.
However, a previous numerical study on a rigid inclusion showed that 
the effect is made negligibly small by 
the hydrodynamics of the ambient 2D and 3D fluids \citep{camley2}.  
It thus becomes of interest whether or not the effect remains negligible when the inclusion is
a circular liquid domain, considering that its fluidity should alter the ambient flow fields definitely. 
In this paper, we show that the effect can become significantly large in practice
for a circular liquid domain by calculating its drag coefficient.   
\\

Our main assumptions are as follows.
A flat fluid membrane is immersed in a 3D one-component fluid
and contains one circular liquid domain.  
The membrane outside the domain is a 2D binary fluid mixture 
lying in the homogeneous phase near the demixing critical point.  The other component
of the membrane is concentrated in the domain, which is sharply
 bounded by the mixture. 
The correlation length of the mixture is much smaller than the
the domain size.  Thus, the domain dynamics can be regarded as independent of
the critical concentration fluctuation\citep{ofk,furu,yabu}, which is remarkable at length scales smaller than the
correlation length.   
The preferential attraction between the domain component and one component of the mixture
is caused by a short-ranged interaction. 
For the mixture, we can apply the hydrodynamics coming from the
free-energy functional coarse-grained up to the correlation length. 
 In the experimental results of \citet{honerBJ, veatch2, honer}, 
the 2D mixture is in the critical regime when
the correlation length is larger than approximately $100\ $nm.
Thus, the Gaussian free-energy functional, which we use in this study, 
is expected to be valid when the correlation length
is much smaller than $100\ $nm (and much larger than the microscopic length).  
In \citet{oldraft}, one of the present authors studied
the drag coefficient of a domain in this situation
by assuming that the preferential attraction is sufficiently weak
and that the membrane viscosities are the same in the domain interior and exterior.
We here calculate the drag coefficient beyond the regime of these assumptions, also correcting errors in
\citet{oldraft}.   \\

Our formulation is shown in Sect.~\ref{sec:form}.
We use the expansion series of the drag coefficient with respect to a dimensionless 
difference between the membrane viscosities inside and outside the domain
and with respect to a dimensionless strength of the preferential attraction. 
The recurrence relations of the expansion coefficients are derived in Sect.~\ref{sec:cp}, with
some details being relegated to Appendix \ref{app:draco}.  
The drag coefficient is calculated in Sect.~\ref{sec:res}, with the numerical procedure being
mentioned in Appendix \ref{app:num}.  Our numerical results are obtained with the aid of Mathematica (Wolfram Research).
Some details on the transport coefficients are mentioned in Appendix \ref{app:mct}, which contains 
extensions of the results of \citet{inafuji}.
Our results are discussed in Sect.~\ref{sec:dis}.

\section{Formulation \label{sec:form}}
\begin{figure}
\begin{center}
		\includegraphics[width=8cm]{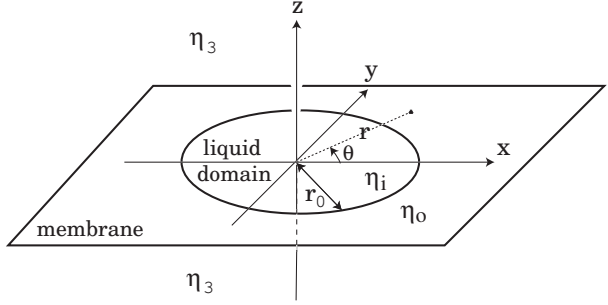}
	\end{center}
	\caption{A circular liquid domain with the radius $r_0$ in a flat fluid membrane, which stretches infinitely over the $xy$-plane. 
	The Cartesian coordinate system $(x,y,z)$ and cylindrical coordinate system $(r,\theta,z)$ are shown. 
	The membrane viscosities are $\eta_{\rm i}$ and $\eta_{\rm o}$ inside and outside the domain, respectively, 
	while the viscosity is $\eta_3$ in the 3D fluid occupying the semi-infinite spaces on both sides of the membrane. 
	}
	\label{incls}
\end{figure}

As shown in Fig.~\ref{incls}, we set the Cartesian coordinate system $(x,y,z)$ and
cylindrical coordinate system $(r,\theta,z)$.  The flat membrane lies on the $xy$ plane.
A circular liquid domain (radius $r_0$) is
fixed with its center being at the origin. The ambient 3D fluid (viscosity $\eta_3$) 
is assumed to occupy the semi-infinite spaces on both sides of the membrane. 
Imposing a weak homogeneous flow far from the domain,
we consider the stationary state and calculate
the total force exerted on the fixed domain in the linear regime.
The quotient of  its magnitude divided by the speed of the homogeneous flow
is the drag coefficient $\gamma$.  \\

In Sect.~\ref{sec:simple}, we assume neither preferential attraction nor near-criticality to 
review the formulation and procedure in \citet{tani}, where the membrane outside the domain
is regarded as a 2D one-component fluid.
In Sect.~\ref{sec:critical}, we mention the points to be altered in the formulation for a near-critical
binary fluid membrane.  The equations in this subsection are
essentially the same as described in \citet{oldraft}.

\subsection{Case of no preferential attraction \label{sec:simple}}
Writing $P$ and ${\bm V}$  
for the pressure and velocity fields
in the 3D fluid, respectively, we have
the Stokes equation and the incompressibility
condition, {\it i.e.\/},
\begin{equation} -\nabla P+\eta_3\Delta {\bm V}=0
\quad {\rm and}\quad  \nabla\cdot{\bm V}=0\label{eqn:3st}
\end{equation} 
for $z\ne 0$.
Writing $p$ and ${\bm v}$ for these fields in the 2D fluid, we similarly have
\begin{equation}
-\nabla p+\eta \Delta {\bm v}+{\bm F}=0
\quad {\rm and}\quad  \nabla\cdot{\bm v}= 0 \label{eqn:2st}
\end{equation}
for $r\ne r_0$ and $z=0$.  
Here, the differential operator is defined in the two dimensions, the membrane 
viscosity $\eta$ equals $\eta_{\rm i}$ inside the domain $(r<r_0)$ and $\eta_{\rm o}$
outside the domain $(r>r_0)$, and
 ${\bm F}$ denotes the stress exerted by the 3D fluid lying on both sides of the membrane.  
Assuming the impermeability of the membrane, we have $V_z\to 0$ as $z\to 0$.
The no-slip condition gives
\begin{equation}
	\lim_{z \to 0} V_r(r, \theta, z) = v_r(r, \theta) \quad {\rm and} \quad \lim_{z \to 0} V_\theta(r, \theta, z) = v_\theta(r, \theta) 
\ . \label{eqn:memnoslip}
\end{equation}
We write $\tau$ for
the stress field of the 2D fluid; its $rr$-component is given by
$\tau_{rr}=-p+2\eta \partial v_r/(\partial r)$.
The stress exerted on the domain perimeter in its tangential direction by the domain interior
should be balanced with the one by the exterior, and thus we have
\begin{equation}
\lim_{r\to r_0+}\tau_{r\theta}=\lim_{r\to r_0-}\tau_{r\theta}\ ,
\label{eqn:tangcond}\end{equation}
where $r\to r_0+\ (-)$ means that $r$ approaches $r_0$ with
$r>r_0\ (<r_0)$ kept.  Thus, $v_\theta$
is not smooth across $r=r_0$ when 
$\eta_{\rm i}$ is not equal to $\eta_{\rm o}$, as mentioned in \citet{confine},
although $V_\theta$ is always smooth in each of the semi-infinite spaces.   \\

The velocity field of the homogeneous flow far from the domain is assumed to be given by 
$-\varepsilon U{\bm e}_x$, where ${\bm e}_x$ is the unit vector along the $x$-axis, 
 $U$ is a nonzero constant with the dimension of velocity, and
$\varepsilon$ is a small dimensionless parameter 
introduced for convenience of later calculations. The total force is along ${\bm e}_x$, and
its $x$-component is given by $-\varepsilon \gamma U$
up to the order of $\varepsilon$. 
Far from the domain,
$p$ and $P$ reach 
constant values, $p^{(0)}_{\rm o}$ and $P^{(0)}$, respectively.
The velocity fields,  ${\bm V}$ and ${\bm v}$, are not equal to
the homogeneous flow everywhere. 
We expand the fields 
with respect to $\varepsilon$.  In the 3D fluid, $P^{(1)}$ and ${\bm V}^{(1)}$ are so defined 
that \begin{equation}
P({\bm r})=P^{(0)}+\varepsilon P^{(1)}({\bm r})\quad {\rm and}\quad 
{\bm V}({\bm r})=-\varepsilon U{\bm e}_x+ \varepsilon {\bm V}^{(1)}({\bm r})
\label{eqn:perexp3}\end{equation}
hold up to the order of $\varepsilon$.  Similarly, 
$p^{(1)}$ and ${\bm v}^{(1)}$ are so defined that
\begin{equation}
p({\bm r})=p^{(0)}+\varepsilon p^{(1)}({\bm r})\quad
{\rm and}\quad {\bm v}({\bm r})=-\varepsilon U{\bm e}_x+ \varepsilon {\bm v}^{(1)}({\bm r})
\label{eqn:perexpmem}\end{equation}
hold up to the order of $\varepsilon$.  Here,
$p^{(0)}$ equals a constant $p_{\rm o}^{(0)}$ outside the domain and
another constant $p_{\rm i}^{(0)}$ inside the domain.
Because the radial component of ${\bm v}$ vanishes 
at the perimeter, we have 
\begin{equation}
v_r^{(1)}=U\cos{\theta}\quad {\rm at}\ r=r_0\ .\label{eqn:vrU}
\end{equation} 
We also define ${\bm F}^{(1)}$ and $\tau^{(1)}$ to have
${\bm F}=\varepsilon {\bm F}^{(1)}$ and $\tau=\varepsilon \tau^{(1)}$
up to the order of $\varepsilon$. 
The fields with the superscript $^{(1)}$ vanishes far from the domain. \\

We introduce the Fourier transforms with respect to $\theta$, {\it e.g.\/},
\begin{equation}
{\tilde V}^{(1)}_{zm}(r, z) \equiv \frac{1}{2\pi} \int^{2\pi}_0 d \theta \ V^{(1)}_z(r, \theta, z) e^{-i m \theta}\ ,
\label{tilVZM}\end{equation}
with $m = 0, \pm 1, \pm2, \dots$.  
Its Hankel transform is given by
\begin{equation}
\int^\infty_0 d r \,\, r J_m(\zeta r)\tilde{V}^{(1)}_{zm}(r, z)\ ,\label{tilVZM2}
\end{equation}
where $J_m$ is the Bessel function of the first kind.
Because of the symmetry, the Fourier transforms with $m\ne \pm 1$ vanish.
In each field, the transforms of $m=\pm 1$ are related with each other.  
Thus, we have only to consider the Fourier transforms with $m=1$.  
As shown in Appendix A of \citet{tani}, we rewrite
Eq.~(\ref{eqn:3st}) into the Hankel transforms 
and solve the resultant ordinary differential equations with
two functions of $\zeta$ being left undetermined.  Here, as in Eq.~(\ref{tilVZM2}),
$\zeta$ is the variable introduced at the Hankel transformation.
We can substitute the solution into Eq.~(\ref{eqn:2st}) with the aid of
Eq.~(\ref{eqn:memnoslip}) to fix the undetermined functions.
As mentioned in Appendix A of \citet{tani},
we can use the incompressibility conditions to
delete one of the two undetermined functions of $\zeta$,
and thus have only to consider one undetermined function of $\zeta$. We write $A(\zeta)$
for this function.   Introducing 
\begin{equation}
\nu_{\rm o} \equiv {\eta_{\rm o}\over 2\eta_3 r_0}\quad {\rm and}\quad
J_{\pm}(\zeta)\equiv J_2(\zeta)\pm J_0(\zeta)\ ,
\end{equation}
we obtain
\begin{equation}
{\tilde V}^{(1)}_{r1} (r, z)= 
{1\over 8\eta_3 {r_0}^2 } \int_0^\infty d\zeta\ {A(\zeta)J_+(\zeta R)\over  1+\nu_{\rm o}\zeta}  e^{-\zeta Z} \ ,
\label{eqn:Vr}\end{equation}
where $R$ and $Z$ are respectively defined as $r/r_0$ and $z/r_0$; 
$i{\tilde V}^{(1)}_{\theta 1}(r, z)$ is given by the above with $J_+$ being replaced by $J_-$. 
The integral above comes from the inverse Hankel transformation.
Because $V_z({\bm r})$ vanishes, we have
\begin{equation}
F^{(1)}_r = 2\eta_3 \lim_{z \to 0+} \frac{\partial V^{(1)}_r}{\partial z} \quad {\rm and} \ 
F^{(1)}_\theta = 2\eta_3 \lim_{z \to 0+} \frac{\partial V^{(1)}_\theta}{\partial z}\ , \label{Frtheta}
\end{equation}
where the factor $2$ comes because the force is exerted from both sides of the membrane. \\

We introduce a dimensionless parameter, defined as \begin{equation}
\kappa\equiv 1-{\eta_{\rm o}\over \eta_{\rm i}}
\ ,\end{equation}
which vanishes for $\eta_{\rm i}=\eta_{\rm o}$.  The domain is regarded as 
a rigid disk when $\kappa$ approaches unity from below.
Extracting the $r$ and $\theta$-components with the order of $\varepsilon$
from the first equation of Eq.~(\ref{eqn:2st}), 
we  substitute 
${\tilde V}^{(1)}_{r1}$ and ${\tilde V}^{(1)}_{\theta 1}$, which are
expressed in terms of $A$, into the Fourier transforms of the components
with the aid of Eq.~(\ref{eqn:memnoslip}).  
Deleting ${\tilde p}_1^{(1)}$ from the results, we arrive at
\begin{eqnarray}&&
\int_0^\infty d\zeta\ \zeta^2 J_1(R\zeta)A(\zeta) =0
\quad {\rm for}\ R>1 \quad {\rm and}\ \label{eqn:calAout0}\\
&&\int_0^\infty d\zeta\ \zeta^2 J_1(R\zeta)A(\zeta) \left(1-{\kappa\over 1+\nu_{\rm o}\zeta}\right)=0
\quad {\rm for}\ 0\le R<1 \label{eqn:calAin}\ ,
\end{eqnarray} 
which are equivalent to Eqs.~(3.5) and (3.6) of \citet{tani}.
When $\kappa$ vanishes, {\it i.e.\/}, when $\eta_{\rm i}$ equals $\eta_{\rm o}$, 
the left-hand side (lhs) of Eq.~(\ref{eqn:calAin}) becomes identical with
that of Eq.~(\ref{eqn:calAout0}).   \\

The $x$-component of the integral of ${\bm F}^{(1)}$ over the domain is given by
\begin{equation}
	\int^{r_0}_0{d r} \, r \, \int^{2\pi}_0 {d \theta} \, \left(F^{(1)}_r \cos{\theta} - F^{(1)}_\theta \sin{\theta}\right) 
	= 2\pi \int^{r_0}_0 dr\ r \left({\tilde F}^{(1)}_{r1} - {\rm i} {\tilde F}^{(1)}_{\theta 1}\right) \label{eqn:EFU}\ .\end{equation}
That of the force exerted on the domain by the membrane outside the domain is given by
\begin{equation}
	r_0 \int^{2\pi}_0 d \theta \  \left(\tau^{(1)}_{rr} \cos{\theta} - \tau^{(1)}_{r\theta} \sin{\theta}\right) 
	= 2 \pi r_0 \left({\tilde\tau}^{(1)}_{rr1} - {\rm i} {\tilde \tau}^{(1)}_{r\theta 1}\right)\ , \label{eqn:TAU}
\end{equation}
which is evaluated in the limit of $r\to r_0+$. 
These two equations can be rewritten 
in terms of ${\tilde V}^{(1)}_{r1}$ and ${\tilde V}^{(1)}_{\theta 1}$ with the aid of
Eqs.~(\ref{eqn:memnoslip}) and (\ref{Frtheta})\citep{pdel}.  
The sum of Eqs.~(\ref{eqn:EFU}) and (\ref{eqn:TAU}) equals $-\gamma U$, and
the drag coefficient $\gamma$ is thus found to be given by \citep{prevgam}
\begin{equation}
{\pi\over r_0U}  \lim_{R \to 1+} \lim_{Z \to 0+}
 \int_0^\infty d \zeta \,\, \zeta J_2(\zeta R) A (\zeta) \, e^{-\zeta Z}
\ .\label{eqn:simplegam}
\end{equation}
The operator
$\Theta$ is so defined that $\Theta A$ denotes
the double limits of the integral above.
The convergence factor $e^{-\zeta Z}$ is originally contained in 
${\tilde V}^{(1)}_{r1}$ and ${\tilde V}^{(1)}_{\theta 1}$, as shown in Eq.~(\ref{eqn:Vr}).
It is later found that $A(\zeta)$  for $\kappa\ne 0$ contains a term proportional to $J_0(\zeta)$, and thus
the convergence factor cannot be dropped from Eq.~(\ref{eqn:simplegam}).
Precisely speaking, each integrand of Eqs.~(\ref{eqn:calAout0}) and (\ref{eqn:calAin}) also has this factor, 
as mentioned below Eq.~(3.24) of \citet{confine}.  \\

We consider the case of $\kappa=0$ in this paragraph.
Because the completeness of the Hankel transformation gives
\begin{equation}
\int_0^\infty d\zeta\ \zeta J_1(R\zeta)J_1(R'\zeta)=\delta(R-R')
\ ,\end{equation}
we find $A(\zeta)\propto J_1(\zeta)/\zeta$  from Eqs.~(\ref{eqn:calAout0}) and (\ref{eqn:calAin}) for $\kappa=0$.  
The constant of proportionality is fixed by Eq.~(\ref{eqn:vrU}),
as mentioned in \citet{confine}.   Thus, for $\kappa=0$,
$A(\zeta)$ equals
\begin{equation}
 {2\eta_3r_0^2 U\over \chi} \times {J_1(\zeta)\over\zeta}\label{eqn:azero}\ ,
\end{equation}
which satisfies Eq.~(\ref{eqn:tangcond}) automatically.
Here, we use \citep{comchi}
\begin{equation}
\chi\equiv \int_0^\infty d\zeta {J_1(\zeta)^2 \over \zeta^2\left(1+\nu_{\rm o}\zeta\right)}\ .
\label{eqn:chidef}
\end{equation}
Substituting Eq.~(\ref{eqn:azero}) into Eq.~(\ref{eqn:simplegam}), we find 
$\gamma=2\pi \eta_3 r_0/\chi$ for $\kappa=0$, which was first obtained in \citet{koker}.  \\

Assuming that $\kappa$ does not always vanish, we introduce
\begin{equation}
{\hat A}(\zeta)\equiv {\chi \over 2\eta_3r_0^2 U }  A(\zeta) \ .\label{eqn:hatadef}
\end{equation}
Noting Eq.~(\ref{eqn:azero}), ${\hat A}(\zeta)$ equals $J_1(\zeta)/\zeta$ when $\kappa$ vanishes.  
For this function, we write ${\hat A}_0^{(0)}(\zeta)$, {\it i.e.\/},
\begin{equation}
{\hat A}^{(0)}_0(\zeta)\equiv  {J_1(\zeta)\over \zeta}\ ,
\end{equation}  where the subscript $_0$ indicates $\kappa=0$ and
the superscript $^{(0)}$ indicates the absence of the preferential attraction.  
It is to be noted that the meaning of this
superscript is different from the one used in Eqs.~(\ref{eqn:perexp3}) and (\ref{eqn:perexpmem}), where
the superscript is added to a field.
The following procedure applicable for $\kappa\ne 0$ is devised in \citet{confine}
and mentioned more explicitly around  Eq.~(3.7) of \citet{tani}.
Irrespective of the value of $R$, we define
 $q(R)$ as the integral of Eq.~(\ref{eqn:calAout0}) with 
$A$ being replaced by ${\hat A}$, {\it i.e.\/},
\begin{equation}
q(R)\equiv \int_0^\infty d\zeta\ \zeta^2 J_1(R\zeta){\hat A}(\zeta){\quad} {\rm for}\ R\ge 0
\ ,\end{equation}
and then define a finite function $q_1(R)$ to have
\begin{equation}
  q(R)=q_1(R)+ c_1 \delta\left(R-1\right) +c_2 {d\over dR} \delta\left(R-1\right)
\ .\label{eqn:qR}
\end{equation}
Here, $c_1$ and $c_2$ are constants independent of $R$, and 
$q_1(R)$ vanishes for $R>1$ because of Eq.~(\ref{eqn:calAout0}).   
The Hankel transformation of Eq.~(\ref{eqn:qR}) involves the integral of $Rq_1(R)J_1(\zeta R)$ over $0<R<1$.
Rewriting this integral with the aid of Eq.~(\ref{eqn:calAin}), 
we arrive at a single integral equation
\begin{equation}
{\hat A} =\kappa \left[{\cal M} {\hat A}\right]
+\left(c_1-\kappa {\cal I}{\hat A}  \right) {\hat A}^{(0)}_0-c_2J_0
\ ,\label{eqn:Atilexpress0}\end{equation}
where
the operators ${\cal M}$ and  ${\cal I}$ are so defined that we have
\begin{eqnarray}&&
[{\cal M}{\hat A}](\xi)=\int_0^\infty d\zeta \ 
 \frac{\xi J_0(\zeta)J_1(\xi)-\zeta J_0(\xi)
J_1(\zeta)}{\xi^2-{\zeta}^2} \times \frac{\zeta {\hat A}(\zeta)}{1+ \nu_{\rm o}\zeta} \label{eqn:Mdef}
\\ && {\rm and}\quad 
{\cal I}{\hat A}=
\int_0^\infty d\zeta\ {\zeta J_0(\zeta){\hat A}(\zeta)
 \over 1+\nu_{\rm o} \zeta}
\ .\label{eqn:Idef}\end{eqnarray}
For $\kappa=0$, we have $q_1(R)\equiv 0$, 
$c_1=1$, $c_2=0$, and Eq.~(\ref{eqn:azero}).
For $\kappa\ne 0$, we should fix $c_1$ and $c_2$ by using the two conditions of 
Eqs.~(\ref{eqn:tangcond}) and (\ref{eqn:vrU}), as described below. In other words, we
generally need two constants to fulfill the two conditions. \\

Let us rewrite the two conditions into convenient forms.
With the aid of Eq.~(\ref{eqn:memnoslip}), Eq.~(\ref{eqn:Vr}) gives
\begin{equation} 
{\tilde v}_{r1}^{(1)}(r)= {U\over 2\chi} \left[{\cal L} {\hat A}\right](R) 
\ ,\label{eqn:vr1}\end{equation} where
the operator ${\cal L}$ is so defined that we have 
\begin{equation}
\left[{\cal L} {\hat A}\right](R) = {1\over R} \int_{0}^\infty d \zeta {J_1(\zeta R){\hat A}(\zeta)\over \zeta\left(
1+\nu_{\rm o}\zeta\right)} 
\ .\label{eqn:Ldef}\end{equation} 
We can rewrite Eq.~(\ref{eqn:vrU}) as 
\begin{equation}
  \left[{\cal L}{\hat A}\right](1)=\chi\ ,\label{eqn:vrop}
\end{equation}
into which Eq.~(\ref{eqn:Atilexpress0}) is substituted to give
\begin{equation}
\kappa\left[{\cal L}{\cal M}{\hat A}\right](1)
+\left( c_1- \kappa {\cal I}{\hat A} \right) \chi-c_2 \left[{\cal L}J_0\right](1)=\chi
\ .\label{eqn:vrU220}\end{equation}
In passing, we have Eq.~(\ref{eqn:chidef}) because 
Eq.~(\ref{eqn:vrop}) holds even for $\kappa=0$, and thus
$\left[ {\cal L}\left({\hat A}-{\hat A}_0^{(0)}\right)\right](1)$ vanishes.
With the aid of Eq.~(\ref{eqn:memnoslip}), Eq.~(\ref{eqn:Vr}) gives
\begin{equation}
{\tilde \tau_{r\theta 1}}={-i \eta U \over 2r_0 \chi} \left[{\cal N}{\hat A}\right](R)\ ,
\label{eqn:tangcondop}\end{equation}
where the operator ${\cal N}$ is so defined that we have
\begin{equation}
\left[{\cal N}{\hat A}\right](R)
= \int^\infty_0 d \zeta\ \frac{\zeta J'_2(\zeta R)}{1+ \nu_{\rm o} \zeta} {\hat A}(\zeta)
\label{eqn:Ndef}\end{equation}
with $J'_2(\zeta)$ denoting $dJ_2(\zeta)/(d\zeta)$.  The prime indicates the derivative.
As shown by Eq.~(3.28) of \citet{confine}, 
$\left[{\cal N} J_0\right](R)$ jumps by $1/\nu_{\rm o}$ as $R$ increases across unity.
We define $g$ as $\left[{\cal N} J_0\right](R)$ in the limit of $R\to 1+$ \citep{gmethod}.
Substituting Eq.~(\ref{eqn:Atilexpress0}) into Eq.~(\ref{eqn:tangcondop}), we find Eq.~(\ref{eqn:tangcond}) to give
\begin{equation}
-\kappa^2 \left[{\cal N}{\cal M}{\hat A}\right](1) 
-\kappa\left(c_1- \kappa{\cal I}{\hat A}  \right)\left[{\cal N}{\hat A}_0^{(0)}\right](1) 
+c_2\left( \kappa g -{1\over \nu_{\rm o}} \right)=0\ .
\label{eqn:tangcond20}\end{equation}
Equations (\ref{eqn:Atilexpress0}), (\ref{eqn:vrU220}), and (\ref{eqn:tangcond20})
determine $c_1$, $c_2$, and ${\hat A}$, and then $\gamma$
with the aid of Eqs.~(\ref{eqn:simplegam}) and (\ref{eqn:hatadef}).

\subsection{Near-critical 2D fluid mixture \label{sec:critical}}
We here assume that the membrane outside the domain is a 2D binary fluid mixture, as mentioned in
the fourth paragraph of Sect.~\ref{sec:intro}.
 The difference in the mass per unit area between the two components of the mixture
can depend on the position ${\bm r}$ in the membrane outside the domain.
We write $\varphi({\bm r})$ for the difference, which represents the local composition.
As in Refs.~\citet{ofk,wetdrop,oldraft}, 
the $\varphi$-dependent part of
the free-energy functional is assumed to be 
\begin{equation}
\int_{r>r_0}d{\bm r}\ \left(f(\varphi({\bm r}))+{1\over 2}M
\left\vert\nabla\varphi({\bm r})\right\vert^2\right)
+r_0\lim_{r\to r_0+} \int_0^{2\pi} d\theta\ f_{{\rm s}}(\varphi({\bm r}))
\ .\label{eqn:glw}\end{equation}
The first integral is the area integral over the membrane outside the domain; $f$ is a
quadratic function,
$M$ is a positive constant, and $\nabla$ represents the two dimensional
gradient. The preferential attraction is represented by
the second term, which is the line integral over the domain perimeter.
The function $f_{\rm s}$ is here assumed to be a linear function \citep{Cahn}.
We write $h$ for the surface field, which is a constant defined as
\begin{equation}
h\equiv -f_{\rm s}'=-{d\over d\varphi}f_{\rm s}(\varphi)\ .
\end{equation}
This amounts to considering the dependence of the line tension
on $\varphi$ very near the domain.  What is characteristic here is that
this dependence causes a significant gradient of $\varphi$ around
the domain in the near-critical 2D fluid. 
Far from the domain, the mixture 
is in the homogeneous phase near the 
demixing critical point.  There, $\varphi$ takes a constant value, for which we write $\varphi_\infty$, and
the chemical potential conjugate to $\varphi$ is given by $\mu^{(0)}\equiv f'(\varphi_\infty)$. 
We thus have
\begin{equation}
f(\varphi)={m\over 2} \left(\varphi-\varphi_\infty\right)^2
+\mu^{(0)}\left(\varphi-\varphi_\infty\right)
\ ,\label{eqn:gauss}\end{equation}
where $m$ is a positive constant proportional to the
temperature measured from the critical one. 
The correlation length is given by $\sqrt{M/m}$. 
We define the dimensionless correlation length as
\begin{equation}
s_{\rm c}\equiv {1\over r_0}\sqrt{{M\over m}}
\  .\end{equation}
\medskip

At the equilibrium without the imposed flow, the chemical potential is homogeneously given by
$\mu^{(0)}$, and
$\varphi$ minimizes the grand-potential functional coming from 
Eq.~(\ref{eqn:glw}).  The minimization yields     
\begin{equation}
\mu^{(0)}=f'(\varphi({\bm r}))
-M\Delta\varphi({\bm r}) \ ,
\label{eqn:hatmudef}
\end{equation} 
and  \begin{equation}
 M{\bm e}_r\cdot \nabla \varphi =-h \quad {\rm at}\ r=r_0+
\label{eqn:phisurface}\ ,\end{equation}
where ${\bm e}_r$ denotes the radial unit vector.  
The equilibrium profile of $\varphi$ depends only on $r$ because of the symmetry, and is thus
 denoted by $\varphi^{(0)}(r)$, which is given by\citep{oldraft}
\begin{equation}
\varphi^{(0)}(r)
=\varphi_{\infty} + 
{hr_0 s_{\rm c} K_0(R s_{\rm c}^{-1}) \over MK_1(s_{\rm c}^{-1})}
\quad {\rm for}\ r>r_0\ ,\label{eqn:phizero}\end{equation} 
where $K_0$ and $K_1$ are modified Bessel functions.
The thickness of the adsorption layer, where the preferred component is 
significantly concentrated, can be regarded as given by the correlation length. 
  For later convenience, we introduce
\begin{equation}
\Phi(R)\equiv 
{K_1(Rs_{\rm c}^{-1})\over K_1(s_{\rm c}^{-1})}
\ ,\label{eqn:kappadef}
\end{equation}
which leads to ${\varphi^{(0)}}'(r)=-h\Phi(R)/M$ for $r>r_0$.   \\

In the stationary state under the imposed flow, $\varphi({\bm r})$ deviates from $\varphi^{(0)}(r)$, and
the chemical potential deviates from $\mu^{(0)}$ to become dependent on ${\bm r}$.
As discussed in Appendix A of \citet{ofk}, the chemical potential, denoted by
$\mu({\bm r})$, remains given by the right-hand side (rhs) of Eq.~(\ref{eqn:hatmudef}).  This 
can be explained by the local equilibrium, which also makes
Eq.~(\ref{eqn:phisurface}) valid in the dynamics \citep{jans, ofk, preprint}.
We assume the transport coefficients outside the domain to be constants independent of the local composition, 
as in Refs.~\citet{ofk, wetdrop, furu, camley2}. 
The diffusive flux between the two components is given by $-L\nabla\mu$, 
where the Onsager coefficient $L$ is a positive constant.  
The mass conservation of each component in the stationary state is represented by   
\begin{equation}
L\Delta \mu ={\bm v}\cdot \nabla\varphi \quad {\rm for}\  r>r_0\ .
\label{eqn:phidyn}\end{equation}
The diffusive flux does not pass
across the perimeter, {\it i.e.\/},
\begin{equation}
{\bm e}_r\cdot \nabla\mu =0 \quad {\rm at} \ r=r_0+\ .\label{eqn:musurface}
\end{equation}
As in Eq.~(\ref{eqn:perexpmem}), 
 $\varphi^{(1)}$ and $\mu^{(1)}$ are so defined that
\begin{equation}
\varphi({\bm r})=\varphi^{(0)}(r)+\varepsilon 
\varphi^{(1)}({\bm r})\ {\rm and}\ 
\ \mu({\bm r})=\mu^{(0)}+\varepsilon \mu^{(1)}({\bm r})\ ,
\label{eqn:perexpout}\end{equation}
hold up to the order of $\varepsilon$. 
The pressure tensor coming from the first term of Eq.~(\ref{eqn:glw}), denoted by $\Pi$, 
is used in the model H \citep{Hohenberg, Onukibook}.  We have
\begin{equation}
\Pi=\left(-f+\mu\varphi-{M\over 2}\left|\nabla\varphi\right|^2 \right){\bm 1}+M\nabla\varphi\nabla\varphi
\ ,\label{eqn:pidef}\end{equation}
where ${\bm 1}$ denotes the isotropic tensor.
The osmotic pressure, $\varphi f'-f$, is thus contained 
in Eq.~(\ref{eqn:pidef}).  The term, $-\nabla\cdot \Pi=-\varphi \nabla\mu$
should be added to the lhs of
the first equation of  Eq.~(\ref{eqn:2st}) for $r>r_0$.  
Accordingly, Eq.~(\ref{eqn:calAout0}) is replaced by
\begin{equation}
\int_0^\infty d\zeta\ \zeta^2 J_1(R\zeta)A(\zeta) =
{2r_0^3 {\varphi^{(0)}}'(r){\tilde \mu}_1^{(1)}(r)\over R}
\quad {\rm for}\ R>1 \ ,\label{eqn:calAout}\\
\end{equation}
while Eq.~(\ref{eqn:calAin}) is unchanged. 
We here introduce 
a dimensionless surface field
\begin{equation}
\lambda \equiv {hr_0 \over M}\sqrt{ {r_0\over 2\eta_3 L}}\ .
\label{eqn:barhdef}\end{equation}
Picking up the terms with the order of $\varepsilon$ from Eq.~(\ref{eqn:phidyn})
and solving the resultant equation formally, 
we can express  ${\tilde \mu}_1^{(1)}$ in terms of Eq.~(\ref{eqn:vr1}), as shown in Appendix \ref{app:draco}.
As a result, we can rewrite Eq.~(\ref{eqn:calAout}) as
\begin{equation}
\int_0^\infty d\zeta\ \zeta^2 J_1(R\zeta){\hat A}(\zeta) =
\lambda^2 { \Phi(R) \over R} \left[\Delta_1^{-1}\Phi ({\cal L} {\hat A}- \chi)\right](R)
\quad {\rm for}\ R>1 \ .\label{eqn:calAout2}
\end{equation}
Here, the operator 
$\Delta_1^{-1}$ is so defined that a function $\Omega$ is transformed into 
\begin{equation}
\left[{\Delta}_1^{-1}\Omega \right]\hspace{-0.5mm}(R) 
=\int_1^\infty dR'\ {G(R,R')\over R}  \Omega(R')
\ ,\label{eqn:delinvdef}\end{equation}
where the kernel $G(R,R')$ is defined as $-(1+{R'}^2)/2$ for $R'\le R$ and 
as $-(1+R^2)/2$ for $1<R<R'$. \\

From Eqs.~(\ref{eqn:calAin}) and (\ref{eqn:calAout2}), we can obtain
a single integral equation in the presence of the preferential attraction in the near-criticality
by using essentially the same procedure
as mentioned in Sect.~\ref{sec:simple}.
Here, for $R>1$, $q_1(R)$ of Eq.~(\ref{eqn:qR}) does not vanish but equals 
the rhs of Eq.~(\ref{eqn:calAout2}).   Accordingly, the term
\begin{equation}
\lambda^2 \left[{\cal H}\Delta_1^{-1}\Phi ({\cal L} {\hat A}- \chi)\right](\zeta) \label{eqn:add}
\end{equation}
is added on the rhs of Eq.~(\ref{eqn:Atilexpress0}).  Here, the operator ${\cal H}$ is so defined that 
a function $\Omega$ is transformed into
\begin{equation}
\left[ {\cal H} \Omega \right](\zeta)={1\over \zeta}  \int_1^\infty dR\ \Phi(R)J_1(\zeta R) \Omega(R)
\label{eqn:Hdef}\ .\end{equation} 
Equation (\ref{eqn:tangcond}) remains valid in the presence of the preferential attraction 
because the free-energy functional given by Eq.~(\ref{eqn:glw}) does not contribute to
the tangential stress exerted on the domain \citep{oldraft}. The same situation for a 3D droplet
is discussed in Appendix D of \citet{wetvisc}.  Equation (\ref{eqn:vrU}) 
also remains available here.  Thus, as in Sect.~\ref{sec:simple}, we can use Eqs.~(\ref{eqn:tangcond}) 
and (\ref{eqn:vrU}) to fix $c_1$ and $c_2$, which are contained in Eq.~(\ref{eqn:Atilexpress0})
supplemented with Eq.~(\ref{eqn:add}). \\

In the presence of the preferential attraction, as in its absence,
${\bm F}$ and $\tau$ still contribute to the drag force.
In calculating the latter contribution, we should note that
the term $-\varphi\nabla\mu$ is added to
the lhs of the first equation of  Eq.~(\ref{eqn:2st}).
Furthermore, the drag force has a contribution from $\Pi$, {\it i.e.\/}, should have a term of
\begin{equation}
-r_0 \lim_{r\to r_0+} \int_0^{2\pi} d\theta\  \Pi\cdot {\bm e}_r\ .
\end{equation}
The contribution
from the second term of Eq.~(\ref{eqn:glw}) vanishes.
Summing up all these contributions, we find that 
$\gamma$ is given by the sum of Eq.~(\ref{eqn:simplegam}) and
\begin{equation}
{2\pi\eta_3 r_0 \lambda^2\over \chi} \int_1^\infty dR\ R\Phi(R)\left[\Delta_1^{-1}\Phi
\left( {\cal L}{\hat A} -\chi\right) \right](R)
\ .\label{eqn:dragterm2}\end{equation}
See Appendix \ref{app:draco} for some details.

\section{Recurrence relations \label{sec:cp}}
We expand ${\hat A}$ with respect to $\kappa$ as
\begin{equation}
	{\hat A}(\zeta) = \sum_{n = 0}^\infty  {\hat A}_n(\zeta)\kappa^n\ , \label{tilan}
\end{equation}
whereby ${\hat A}_n(\zeta)$ is defined.
Similarly, we expand $c_1$ and $c_2$ as
\begin{equation}
c_1=\sum_{n=0}^\infty \alpha_n^\sharp \kappa^n\quad {\rm and}
\quad c_2=-\sum_{n=0}^\infty \beta_n \kappa^n\ ,\label{eqn:expcoef}
\end{equation}
where $\alpha^\sharp_n$ and $\beta_n$ are the expansion coefficients independent of $\kappa$.
Then, we define $\alpha_n$ as $\alpha^\sharp_n-{\cal I} {\hat A}_{n-1}$ for $n\ge 1$ with $\alpha_0$ being put equal to $\alpha_0^\sharp$.
The dynamics is invariant against the exchange of the components in the 2D mixture, and thus
against the change of the sign of $h$. 
Thus, we can expand ${\hat A}_n$, $\alpha_n$, and $\beta_n$ with respect to $\lambda^2$,
which represents the strength of the preferential attraction, as
\begin{equation}
{\hat A}_n=\sum_{k=0}^\infty  {\hat A}_n^{(k)}\lambda^{2k}\ ,\ 
\alpha_n=\sum_{k=0}^\infty  \alpha_n^{(k)}\lambda^{2k}\ ,\ {\rm and}\
\beta_n=\sum_{k=0}^\infty  \beta_n^{(k)}\lambda^{2k}\label{eqn:explam}
\ ,\end{equation} whereby $ {\hat A}_n^{(k)}$, $\alpha_n^{(k)}$, and $\beta_n^{(k)}$ are defined.
We similarly expand the drag coefficient as
\begin{equation} \gamma ={2\pi\eta_3 r_0\over \chi}\left(\sum_{n=0}^\infty {\hat \gamma}_n \kappa^n\right)
\quad{\rm and}\quad  {\hat \gamma}_n=\sum_{k=0}^\infty {\hat \gamma}_n^{(k)}\lambda^{2k} 
\ .\label{eqn:expgam}
\end{equation}
The expansion coefficients, ${\hat \gamma}_n$ and ${\hat \gamma}_n^{(k)}$, are here introduced.
We have ${\hat \gamma}_0^{(0)}=1$ because of the statement below Eq.~(\ref{eqn:chidef}).
The sum in the parentheses of the first entry of Eq.~(\ref{eqn:expgam})  
gives the ratio of $\gamma$ to its value for $\kappa=\lambda=0$
and is below referred to as dimensionless drag coefficient, for which we write ${\hat \gamma}$.
From Eqs.~(\ref{eqn:simplegam}) and (\ref{eqn:dragterm2}), we find
\begin{equation}
{\hat \gamma} = \Theta {\hat A}+\lambda^2 
\int_1^\infty dR\  R\Phi(R)\left[\Delta_1^{-1}\Phi
\left( {\cal L}{\hat A} -\chi\right) \right](R) 
\ ,\label{eqn:gam}\end{equation}
where the operator $\Theta$ is defined below Eq.~(\ref{eqn:simplegam}).

\subsection{Terms studied previously \label{sec:prev}}
The terms with $k=0$ in Eqs.~(\ref{eqn:explam}) and (\ref{eqn:expgam}) are calculated
in  \citet{tani}, where the drag coefficient is studied 
in the absence of the preferential attraction, {\it i.e.\/}, for $\lambda=0$.
For convenience of later description, we here review their 
recurrence relations.
Substituting Eqs.~(\ref{tilan}) and (\ref{eqn:expcoef}) into
 Eqs.~(\ref{eqn:Atilexpress0}), ({\ref{eqn:vrU220}), and (\ref{eqn:tangcond20}),  
we find $\alpha_0^{(0)}=1$, $\beta_0^{(0)}=0$,  ${\hat A}_0^{(0)}(\zeta)=J_1(\zeta)/\zeta$, 
 \begin{eqnarray}
 && \beta_n^{(0)} = \nu_{\rm o} \left( \alpha_{n - 1}^{(0)} \left[{\cal N}{\hat A}_0^{(0)}\right](1) +  
\beta_{n - 1}^{(0)} g+\left[{\cal N} {\cal M} {\hat A}_{n - 2}^{(0)}\right](1) 
      \right)  \ ,\label{beta}\\
 &&\alpha_n^{(0)} = - \frac{1}{\chi} \left( \beta_n^{(0)} \left[{\cal L}J_0\right](1)+ 
\left[{\cal L}{\cal M}{\hat A}^{(0)}_{n - 1}\right](1) 
\right)  \ , \label{alpha}    \\     
&& {\rm and} \quad
{\hat A}_n^{(0)} = \alpha_n^{(0)} {\hat A}_0^{(0)} + \beta_n^{(0)} J_0 + 
 \left[{\cal M}{\hat A}^{(0)}_{n-1}\right]  \label{hata}
\end{eqnarray}
for $n=1,2,\ldots$, 
where we stipulate ${\hat A}_{-1}^{(0)}\equiv 0$.   
Putting $\lambda$ equal to zero in Eq.~(\ref{eqn:expgam}), we find
the dimensionless drag coefficient to be given by  
\begin{equation}
{\hat \gamma}=1+\sum_{n=1}^\infty {\hat \gamma}_n^{(0)} \kappa^n\quad {\rm for}\ \lambda=0 \ .\label{eqn:gamtani}
\end{equation}
For convenience of numerical calculations, we utilize Eq.~(\ref{hata}) in Eq.~(\ref{eqn:gam}) to obtain
\begin{equation}
	{\hat \gamma}_n^{(0)} =  
\alpha_n^{(0)} + 2\beta_n^{(0)} + \Theta {\cal M} {\hat A}^{(0)}_{n - 1}
\quad {\rm for}\ n=0,1,2,\ldots \ .\label{gamman}
\end{equation}
With the aid of numerical integrations, Eq.~(\ref{eqn:gamtani}) is calculated
for $-1< \kappa < 1$ and some values of $\nu_{\rm o}$ in \citet{tani}, 
where the calculation results in the limit of $\kappa\to 1-$
are shown to agree with the results for a rigid disk in Saffman \& Delbr{\" u}ck's model
\citep{saffman, saffman2, hughes}.  Figure \ref{fig:hgzero} shows numerical results of Eq.~(\ref{eqn:gamtani}),
which gives the ratio of $\gamma$ for $\lambda=0$ to its value for $\kappa=\lambda=0$.  As mentioned in \citet{tani},
it is hard to calculate Eq.~(\ref{eqn:gamtani}) numerically for $\kappa<-1$. 

\begin{figure}[t]
\begin{center}
\includegraphics[width=6cm]{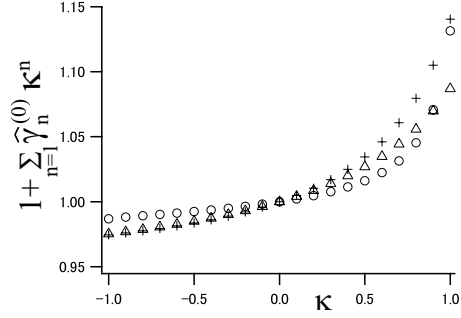}
\end{center}
\caption{Plot of Eq.~(\ref{eqn:gamtani}) against $\kappa$, which is a replot of the results in \citet{tani}.
Circles, crosses, and triangles represent the results
for $\nu_{\rm o}=0.1, 1$, and $10$.  For these values, we
calculate Eq.~(\ref{eqn:gamtani}) by truncating
the series appropriately, {\it i.e.\/}, up to $n=35$, $15$, and $10$, respectively, as in \citet{tani}.}
\label{fig:hgzero}
\end{figure}

\subsection{Higher-order terms with respect to $\lambda^2$ \label{sec:high}}
In the presence of the preferential attraction in the near-criticality, 
substituting Eq.~(\ref{eqn:Atilexpress0}) supplemented with Eq.~(\ref{eqn:add}) into
Eq.~(\ref{eqn:vrop}) yields Eq.~(\ref{eqn:vrU220}) supplemented with  
\begin{equation}
\lambda^2 \left[{\cal L} {\cal H}{ \Delta}_1^{-1} 
\Phi \left( {\cal L}{\hat A}-\chi \right) \right](1) 
\label{eqn:add2}\end{equation}
on the lhs.  Noting that $\left[{\cal N} {\cal H}{\Delta}_1^{-1} 
\Phi\left( {\cal L} {\hat A}-\chi \right) \right](R)$ is continuous at
$R=1$, we similarly find that the term
\begin{eqnarray}
&&-\kappa \lambda^2\left[ {\cal N} {\cal H}{\Delta}_1^{-1} 
\Phi\left( {\cal L} {\hat A}-\chi \right) \right](1)
\label{eqn:add3}\end{eqnarray}
is supplemented on the lhs of
Eq.~(\ref{eqn:tangcond20}).  We use these supplementations to modify the recurrence relations Eqs.~(\ref{beta})-(\ref{hata}),
as mentioned in Appendix \ref{app:draco}, where we also modify Eq.~(\ref{gamman}) by using
Eq.~(\ref{eqn:gam}).  Each of 
${\hat \gamma}_n^{(1)}, {\hat\gamma}_n^{(2)}, \ldots$ depends on $\nu_{\rm o}$ and $s_{\rm c}$.

\section{Results\label{sec:res}}
In the presence of
the preferential attraction in the near-criticality,
we study the drag coefficient of a domain with $\eta_{\rm i}=\eta_{\rm o}$ 
in Sect.~\ref{sec:res1}, and study how the drag coefficient changes with $\kappa\equiv 1-\left(\eta_{\rm o}/\eta_{\rm i}\right)$
in Sect.~\ref{sec:res2}.   What coefficients in the second series of Eq.~(\ref{eqn:expgam}) are discussed
in each subsection is summarized in Fig.~\ref{fig:nktab}.  
The recurrence relations shown in Sect.~\ref{sec:high} and the following results are newly obtained
in the present study.  
Errors in \citet{oldraft}, where only ${\hat\gamma}_0^{(1)}$ was studied, are pointed out in Appendix \ref{app:draco}.

\begin{figure}[t]
\begin{center}
\includegraphics[width=6cm]{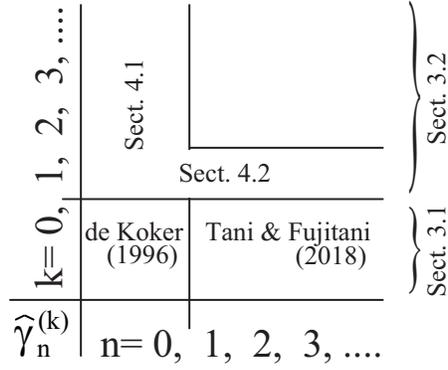}
\end{center}
\caption{The values of $(n,k)$ for ${\hat\gamma}_n^{(k)}$ discussed in each subsection are shown. 
In the absence of the preferential attraction $(\lambda=0)$,
de Koker \citep{koker} found that $\gamma$ for $\kappa=0$
is given by $2\pi\eta_3 r_0/\chi$.  Dividing $\gamma$ by de Koker's
result, we consider the series expansion of the quotient with respect to
$\kappa$ and $\lambda^2$, as shown by Eq.~(\ref{eqn:expgam}); ${\hat\gamma}_0^{(0)}$ equals unity.
The series for $\lambda=0$ is given by Eq.~(\ref{eqn:gamtani})
and was studied by Tani \& Fujitani \citep{tani}, as described in Sect.~\ref{sec:prev}.
The series in the presence of the preferential attraction is mentioned  in Sect.~\ref{sec:high}.    
We show the results of
${\hat\gamma}_0^{(1)}, {\hat\gamma}_0^{(2)}, {\hat\gamma}_0^{(3)}, \ldots$ in Sect.~\ref{sec:res1}, and
those of ${\hat\gamma}_0^{(1)}, {\hat\gamma}_1^{(1)},  {\hat\gamma}_2^{(1)},\ldots$ in Sect.~\ref{sec:res2}.}
\label{fig:nktab}
\end{figure}

\subsection{Results for $\kappa=0$\label{sec:res1}} 
Putting $\kappa$ equal to zero in Eq.~(\ref{eqn:expgam}),
we find the dimensionless drag coefficient
to be given by
\begin{equation}
{\hat\gamma}= {\hat\gamma}_0
= 1+ \sum_{k=1}^\infty {\hat \gamma}_0^{(k)}\lambda^{2k}\quad{\rm for}\ \kappa=0 \ .\label{eqn:res1top}
\end{equation}
The second term on the rhs above 
represents the ratio of the deviation of $\gamma$ from de Koker's result\citep{koker} --
the drag coefficient with $\eta_{\rm i}=\eta_{\rm o}$ in the absence of
the preferential attraction.
We can calculate ${\hat\gamma}_0^{(1)}, {\hat\gamma}_0^{(2)}, \ldots$ by
putting $n$ to be equal zero in the recurrence relations mentioned in Sect.~\ref{sec:high}, {\it i.e.\/},
Eqs.~(\ref{alpha2})-(\ref{gamman1}), where
some terms vanish because of $\beta_0^{(k)}=A_{-1}^{(k)}=0$.
The initial terms $\alpha_0^{(0)}$ and  ${\hat A}_0^{(0)}$ 
are mentioned in Sect.~\ref{sec:prev}.
Our numerical procedure for calculating the integrals contained in the 
expansion coefficients is described in Appendix \ref{app:num}.  
To show their dependence on $\nu_{\rm o}$ and $s_{\rm c}$,
we plot the ratio $\left|{\hat \gamma}_0^{(k)}\right|/s_{\rm c}^{4k}$ against $s_{\rm c}$ for 
$k=1,\ldots,4$ in Fig.~\ref{fig:kdep}.
The results of ${\hat \gamma}_0^{(1)}$ and ${\hat \gamma}_0^{(3)}$ are positive, while 
those of ${\hat \gamma}_0^{(2)}$ and ${\hat \gamma}_0^{(4)}$ are negative.  In
Fig.~\ref{fig:kdep}, $\left|{\hat \gamma}_0^{(k)}\right|$
is smaller than a constant multiplied by $s_{\rm c}^{4k}$, which suggests that 
Eq.~(\ref{eqn:res1top}) converges at least when $\lambda^2s_{\rm c}^4$ is smaller than unity.
As expected, the series appears to converge even for sufficiently large values of $\lambda^2$ in Fig.~\ref{fig:sum},
where ${\hat\gamma}_0^{(k)}$ is positive for $k=1,3,5,\ldots$ and negative for $k=2,4,6,\ldots$.
For $\lambda^2=5000$ in this figure, the results appears independent of $K$ and thus
$1+{\hat\gamma}_0^{(1)}\lambda^2$ is suggested to give a good approximation to
Eq.~(\ref{eqn:res1top}).  It is suggested to give a rather good approximation for $\lambda^2=10^4$.
For larger $\lambda^2$,
some higher-order terms would be required for estimating the sum to the second decimal place.
In our numerical results not shown here, Eq.~(\ref{eqn:res1top}) for 
$(\nu_{\rm o}, s_{\rm c})=(1, 0.1)$ appears to converge even when $\lambda^2$ is raised up to $10^5$ approximately;  
the upper bound of $\lambda^2$ for the convergence increases as $\nu_{\rm o}$ increases. 
This can be expected from Fig.~\ref{fig:kdep};
${\hat\gamma}_0^{(k)}$ for the larger value of $\nu_{\rm o}$
decreases more rapidly as $k$ increases.   In Fig.~\ref{fig:kdep},
$\left|{\hat\gamma}_0^{(k)}\right|$ for a given set of $k$ and $s_{\rm c}$ 
is smaller for the larger value of $\nu_{\rm o}$, and
$\left|{\hat\gamma}_0^{(k)}\right|$ increases with $s_{\rm c}$ for a given set of $k$ and $\nu_{\rm o}$.  Thus,
$\left|{\hat\gamma}_0^{(k)}\right|$ would increase as $s_{\rm c}$ increases and as $\nu_{\rm o}$ decreases. 
\\

\begin{figure}[t]
\begin{center}
\includegraphics[width=12cm]{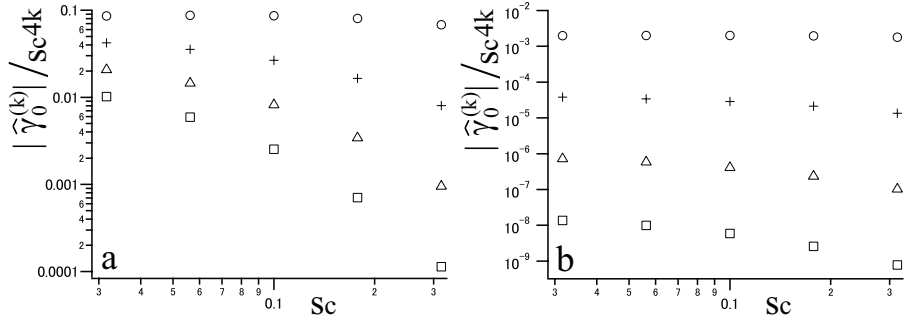}
\end{center}
\caption{Plots of $\left|{\hat \gamma}_0^{(k)}\right|/s_{\rm c}^{4k}$ against $s_{\rm c}$
for $\nu_{\rm o}=0.1$ (a) and $10$ (b).
Circles, crosses, triangles, and squares represent the values for $k=1$, $2$, $3$, and $4$, respectively.}
\label{fig:kdep}
\end{figure}

\begin{figure}[t]
\begin{center}
\includegraphics[width=6cm]{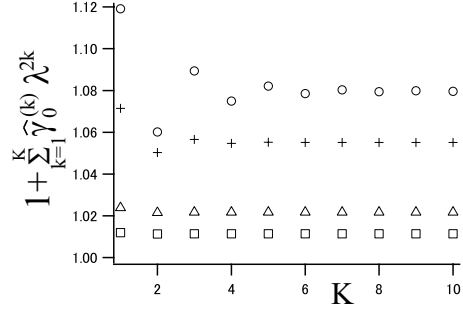}
\end{center}
\caption{Plot of 
the sum of ${\hat \gamma}_0^{(k)}\lambda^{2k}$ form $k=0$ to $K$ against $K$
for $\nu_{\rm o}=1$ and $s_{\rm c}=0.1$.  We use $\lambda^2=50000$ (circles), $30000$ (crosses), $10000$ (triangles), and 
$5000$ (squares). }
\label{fig:sum}
\end{figure}

The Onsager coefficient $L$ of Eq.~(\ref{eqn:phidyn}) depends on the correlation length because of the 
critical fluctuation, as mentioned in Appendix \ref{app:mct}.
The dependence of $\eta_{\rm o}$ on the correlation length is there shown to be very weak and is here neglected.
Noting  Eq.~(\ref{eqn:inafuji0}) and the statement above it, we rewrite $L$ in Eq.~(\ref{eqn:barhdef}) to obtain
\begin{equation}
\lambda^2 \approx {4\pi h^2 r_0^2 
\over M\nu_{\rm o}k_{\rm B}T}   E(N)
\ ,\label{eqn:lamprac}\end{equation}  
where $k_{\rm B}$ is the Boltzmann constant, $T$ is the temperature, $N$ is defined as $s_{\rm c}/\nu_{\rm o}$, and
$E$ is a function defined as
\begin{equation}
E(N)\equiv {2\left(1+N^2\right)\over N^2\left( \pi N-2\ln{N}\right) }\ .
\end{equation}
In Fig.~\ref{fig:E}, $E(N)$ decreases as $N$ increases and
$E(N)$ is approximated to be $-1/N^2/\ln{N}$ for $N\ll 1$.
We can also find from this figure that $NE(N)$ decreases as $N$ increases.
Thus, $\lambda^2$ increases as $\nu_{\rm o}$ increases and as $s_{\rm c}$ decreases. 
\\

\begin{figure}[t]
\begin{center}
\includegraphics[width=6cm]{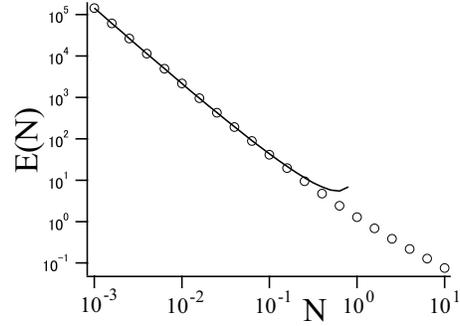}
\end{center}
\caption{Plot of $E(N)$ against $N$ (circles).  The solid curve drawn for $N< 1$ represents $-1/N^2/\ln{N}$. }
\label{fig:E}
\end{figure}

As mentioned in \citet{sphere}, 
when a wall is in contact with a 3D near-critical binary fluid mixture, 
a typical value of the surface field is $10^{-7}$ m$^3$/s$^2$ on the basis of the discussion in \citet{liu},
while one of the coefficient of the square gradient term in the free-energy functional is
$10^{-16}$ m$^7$/(s$^2\cdot$kg) \citep{aiche, vdwexp}.
We use the former value for $h$ in the membrane, and divide the latter value by a typical membrane thickness, {\it i.e.\/},
several nanometers, to use the quotient for $M$ in the membrane, {\it i.e.\/}, to use $M=10^{-8}$ m$^6$/(s$^2\cdot$kg). 
We assume $r_0=100\ $nm; typical values of $\eta_{\rm o} \approx 10^{-7}$ dyn$\cdot$s/cm \citep{smeu} 
and $\eta_3 \approx 10^{-2}$ dyn$\cdot$s/cm$^2$ yield $\nu_{\rm o}\approx 1$.
Assuming $T=300\ $K, we find the fraction in Eq.~(\ref{eqn:lamprac})
to be $30$ approximately.  We also study cases of $\nu_{\rm o}=0.1$ and $10$, for which
the values of the fraction are put equal to $300$ and $3$, respectively. 
For these parameter values, we numerically 
calculate the ratio of the deviation due to the preferential attraction, which is given by
the second term on the rhs of Eq.~(\ref{eqn:res1top}), by truncating the series up to $k=10$,
and plot the results in Fig.~\ref{fig:scdep}.  As mentioned below, we can obtain
almost the same figure by using only the term with $k=1$ in the series.
The difference between the single term ${\hat\gamma}_0^{(1)}\lambda^{2}$  and the sum of the $10$ terms 
increases as $s_{\rm c}$ increases and as $\nu_{\rm o}$ decreases, but rather insensitive to $\nu_{\rm o}$. 
The differences at $\nu_{\rm o}=0.1$ 
are respectively $4\times 10^{-5}$, $5\times 10^{-4}$, and
$5\times 10^{-3}$ for $s_{\rm c}=0.1$, $0.18$, and $0.32$.  
They are respectively much smaller than the corresponding results in  Fig.~\ref{fig:scdep}.
\\

\begin{figure}[t]
\begin{center}
\includegraphics[width=6cm]{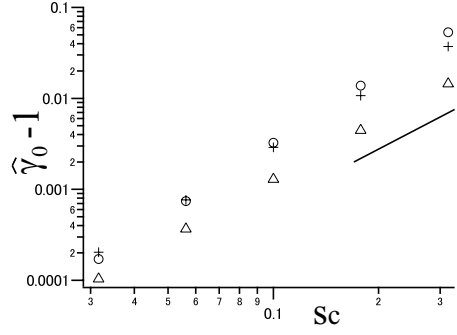}
\end{center}
\caption{Plots of ${\hat \gamma}_0-1$ against $s_{\rm c}$
for $\nu_{\rm o}=0.1$ (circles), $1$ (crosses), and $10$ (triangles).  We calculate the second term on the
rhs of Eq.~(\ref{eqn:res1top}) by truncating the series up to $k=10$.
The values of $\lambda^2$ are obtained from Eq.~(\ref{eqn:lamprac}) and the parameter values mentioned in the text,
They are respectively
$10^3$ multiplied by $2.0$, $0.86$, $0.38$, $0.18$, and $0.086$ for the circles from the extreme left to the right, 
$10^3$ multiplied by $8.6$, $3.2$, $1.2$, $0.49$, and $0.20$ for the crosses, and
$10^3$ multiplied by $52$, $18$, $6.5$, $2.3$, and $0.86$ for the triangles.
The solid line with the slope of two is drawn for a guide of view.}
\label{fig:scdep}
\end{figure}

In Fig.~\ref{fig:scdep}, the ratio of the deviation due to the preferential attraction
increases with $s_{\rm c}$ and is roughly proportional to $s_{\rm c}^2$;
the ratio reaches several percent for $\nu_{\rm o}=0.1$ and $1$.  
The increase with $s_{\rm c}$ is reasonable, considering that the adsorption layer becomes thicker 
with $s_{\rm c}$.   As mentioned above, the increase of 
${\hat\gamma}_0^{(1)}$ and the decrease of $\lambda^2$ occur
when $s_{\rm c}$ increases and when $\nu_{\rm o}$ decreases.
The increase of the deviation ratio with $s_{\rm c}$ shown in Fig.~\ref{fig:scdep}
represents the predominance of the effect of the increase of  ${\hat\gamma}_0^{(1)}$ over that of the decrease of $\lambda^2$.
This predominance can also explain that the deviation ratio for a given $s_{\rm c}$ is the smallest  for $\nu_{\rm 0}=10$.
For larger $\nu_{\rm o}$, effects of the viscous stress in the 2D mixture increases and
the relative contribution from $\Pi$ to the drag force would decrease.
However, the deviation ratio is not so much different for $\nu_{\rm o}=1$ and $0.1$ in Fig.~\ref{fig:scdep}, which suggests
that the effects of ${\hat\gamma}_0^{(1)}$ and $\lambda^2$ are balanced with each other. 
As described in the caption of Fig.~\ref{fig:scdep}, $\lambda^2$ is approximately $1200$ for $\nu_{\rm o}=1$
and $s_{\rm c}=0.1$.  If  $h$ doubles and $M$ halves from their respective estimates mentioned above, $\lambda^2$
increases up to $10^4$ approximately because of
Eq.~(\ref{eqn:lamprac}).   Then,  
the deviation ratio 
becomes approximately $2\ \%$ in Fig.~\ref{fig:sum}, where the ratio is found to
reach several percent when $\lambda^2$ doubles or triples further.  

\subsection{Results up to the order of $\lambda^2$\label{sec:res2}} 
Because of Eq.~(\ref{eqn:expgam}), we have 
\begin{equation}
{\hat\gamma}=1+\sum_{n=1}^\infty {\hat\gamma}_n^{(0)} \kappa^n+
\lambda^2 \sum_{n=0}^\infty {\hat \gamma}_n^{(1)}\kappa^n  \label{eqn:res2top}
\end{equation}
up to the order of $\lambda^2$.
The sum of the first two terms on the rhs above is calculated in Sect.~\ref{sec:prev}.
Here, we numerically calculate ${\hat\gamma}_n^{(1)}$ 
by using Eqs.~(\ref{beta2})-(\ref{gamman1}).  
In Fig.~\ref{fig:tani}(a), ${\hat\gamma}_0^{(1)}$
is positive, while some of the subsequent terms, such as ${\hat\gamma}_1^{(1)}, {\hat\gamma}_2^{(1)}$, and
${\hat\gamma}_3^{(1)}$, are negative.  In Fig.~\ref{fig:tani}(b), 
the third term on the rhs of Eq.~(\ref{eqn:res2top}) decreases as $\kappa$ increases
from $-1$ to $1$.  As $\kappa$ approaches unity,
the third term becomes close to zero, which means that 
the sum of the negative terms then almost cancel out the
positive term.  The positivity of ${\hat\gamma}_0^{(1)}$ comes from the term involving $\chi$ in Eq.~(\ref{eqn:gam}),
and this term does not appear explicitly in the expression of ${\hat\gamma}_n^{(k)}$ for $(n,k)\ne (0,1)$,
as mentioned at the end of Appendix \ref{app:draco}. 
In passing, the result  for $\kappa=0$ in Fig.~\ref{fig:tani}(b) is given by $\lambda^2{\hat\gamma}_0^{(1)}$
and agrees well with the value shown by 
the cross for $s_{\rm c}=0.1$ in Fig.~\ref{fig:scdep}, as expected from the results of Fig.~\ref{fig:sum}.
\\

\begin{figure}[t]
\begin{center}
\includegraphics[width=12cm]{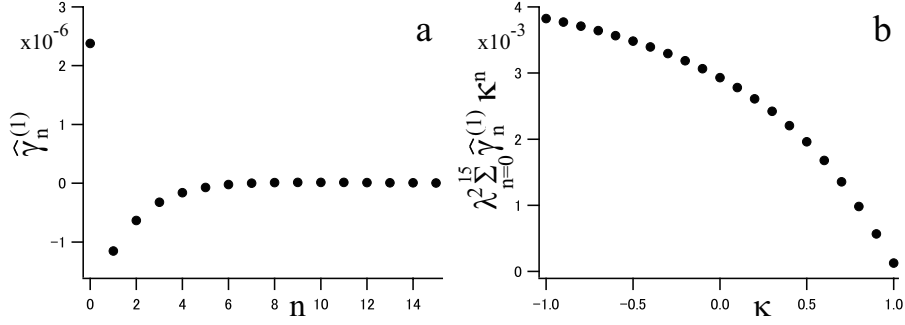}
\end{center}
\caption{Results for $\nu_{\rm o}=1$ and $s_{\rm c}=0.1$.
(a) Plot of ${\hat \gamma}_n^{(1)}$ against $n$. (b)  Plot of the third term on the rhs of Eq.~(\ref{eqn:res2top})
against $\kappa$.  Numerically, we truncate the series given by the third term 
up to $n=15$, like the corresponding series used in Fig.~\ref{fig:hgzero}.
We use $\lambda^2=1.2\times 10^3$, which is 
the same as used for $\nu_{\rm o}=1$ and
$s_{\rm c}=0.1$ in Fig.~\ref{fig:scdep}.
}
\label{fig:tani}
\end{figure}

The ratio of the deviation due to  the preferential attraction is given by
the quotient of the third term divided by the sum of the first two terms on the rhs of Eq.~(\ref{eqn:res2top}).
For $s_{\rm c}=0.1$ and some values of $\nu_{\rm o}$,
we numerically calculate this deviation ratio and plot the results in Fig.~\ref{fig:kappadep}.
The ratio for each of the values of $\nu_{\rm o}$ monotonically decreases to approach zero as $\kappa$ increases to unity,
like the results in Fig.~\ref{fig:tani}(b).  We find that 
the results for $s_{\rm c}=0.1$ in Fig.~\ref{fig:scdep} 
respectively agree well with the corresponding results for $\kappa=0$ in Fig.~\ref{fig:kappadep}.
This means that,
for the parameter values used in this figure, ${\hat\gamma}$ is well given by
Eq.~(\ref{eqn:res2top}) when $\kappa$ vanishes.  


\begin{figure}[t]
\begin{center}
\includegraphics[width=6cm]{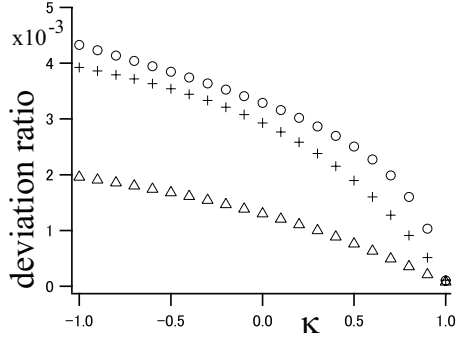}
\end{center}
\caption{Plot of the ratio of the third term to the sum of the first two terms on the rhs of Eq.~(\ref{eqn:res2top})
against $\kappa$ for $s_{\rm c}=0.1$.  Circles, crosses, and triangles represent
the results for  $\nu_{\rm o}=0.1$, $1$, and $10$, respectively.
Numerically, for each value of $\nu_{\rm o}$, the two series in  Eq.~(\ref{eqn:res2top}) are truncated in
the same way as used in Fig.~\ref{fig:hgzero}.
For these values of $\nu_{\rm o}$,  as in Fig.~\ref{fig:scdep},
we respectively use $\lambda^2=3.8\times 10^2, 1.2\times 10^3$, and
$6.5\times 10^3$. }
\label{fig:kappadep}
\end{figure}

\section{Discussion \label{sec:dis}}
We consider a circular liquid domain, which is put in a flat fluid membrane composed of
a binary fluid mixture lying in the homogeneous phase near the demixing critical point. 
The correlation length is assumed to be much smaller than the domain size.
The domain dynamics is thus regarded as independent of the critical fluctuation, which is significant at length scales
smaller than the correlation length and makes the Onsager coefficient dependent on the correlation length. 
 Because of the near-criticality, the preferential attraction between the domain component and one of the component of the mixture
generates the composition gradient outside the domain significantly and
can affect the drag coefficient.   The correlation length is also assumed to be so small
as to validate the Gaussian model.  \\

In Sect.~\ref{sec:res1}, we calculate the dimensionless drag coefficient for $\kappa=0$, {\it i.e.\/}, for
a domain with $\eta_{\rm i}=\eta_{\rm o}$.  
There, the expansion series with respect to
$\lambda^2$, which represents the strength of the preferential attraction, is found to be helpful even for large values of $\lambda^2$
because the expansion coefficient decreases much as the order of $\lambda^2$ increases. 
We can obtain a good approximation to the second term on the rhs of
Eq.~(\ref{eqn:res1top}) by calculating only the first term of the series 
for practical values of $\lambda^2$.   As shown in Fig.~\ref{fig:scdep},
the deviation of the drag coefficient for $\kappa=0$ due to the preferential attraction increases with the correlation length
and can reach several percent of the one in the absence of the preferential attraction
for some practical material constants.  
Dependence of the deviation on the correlation length is steeper than linear. 
The drag coefficient for $\kappa=\lambda=0$ was obtained in \citet{koker} and is given below Eq.~(\ref{eqn:chidef});
its dependence on $r_0$ is not steeper than linear, as shown in Fig.~4 of \citet{yf2011}.
Thus, the dependence of the deviation cannot be explained naively
by effective enlargement of the domain radius due to the adsorption layer, whose thickness
can be regarded as the correlation length.  \\

In Sect.~\ref{sec:res2}, we calculate the dimensionless drag coefficient up to the order of $\lambda^2$, not assuming $\kappa=0$.
In  Fig.~\ref{fig:kappadep}, the effect of the preferential attraction decreases as $\kappa$ increases from $-1$ to $1$ and
becomes negligibly small for a rigid disk.  
The truncation up to the order of $\lambda^2$ gives a good approximation for the parameter values used
in this figure when $\kappa$ vanishes.  We expect that it remains the case for $\kappa\ne 0$, considering
that, as far as examined, the magnitudes of the expansion coefficients are correlated with each other 
through the recurrence relations.  
Our results qualitatively agree with 
one of the results of \citet{camley2} showing that the effect on a rigid disk should be negligibly small.
The results were obtained with the aid of the immersed boundary method\citep{peskin}, where
the boundary condition at the perimeter is altered from the no-slip condition.  In this previous study, its was also claimed that
the hydrodynamics reduces the effect on a rigid disk.  This is also consistent with our results
showing that the effect becomes larger for  a less viscous liquid domain, whose fluidity should alter the
flow fields more definitely from the one around a rigid disk with the same size.  \\

For detecting the effect of the preferential attraction by measuring how $\gamma$ changes with the correlation length,
$\nu_{\rm o}\stackrel{<}{\sim} 1$ and smaller $\kappa$ are suggested to be favorable in Figs.~\ref{fig:scdep}
and  \ref{fig:kappadep}.
The expansion coefficients ${\hat\gamma}_n^{(k)}$ for $n=1,2,\ldots$ and $k=2,3,\ldots$
are not calculated in the present study, as shown in Fig.~\ref{fig:nktab}.
Various integrals of damped-oscillating functions
over a semi-infinite intervals are involved in the recurrence relations, as shown in Appendix \ref{app:num}.
We need to improve the procedure to shorten the computing time for the purpose of calculating 
the coefficients extensively.  
The deviation ratios shown in Fig.~\ref{fig:kappadep} almost vanish for a rigid disk irrespective of 
the values of $\nu_{\rm o}$.  However, the ratio for a rigid disk
may be definitely positive or negative for some parameter values of $(\nu_{\rm o}, s_{\rm c})$ not yet
examined.  This point remains to be studied.
Calculating the drag coefficient of a domain
beyond the regime of the Gaussian model is another future work.  We would have to
consider the inhomogeneity of the correlation length and transport coefficients
beyond the regime, judging from the 3D mixture in a similar situation \citep{preprint}.
Some clues may be obtained from 
studies on the static properties of the membrane
containing domains\citep{mach, nowa}.

\acknowledgements{ 
A part of the work was financially
supported by  Keio Gakuji Shinko Shikin.}

\appendix
\section{Some details \label{app:draco}}
For $R>1$, Eq.~(\ref{eqn:phidyn}) gives
\begin{equation}
 \left({\partial^2\over\partial R^2}+{1\over R}
{\partial\over\partial R}
 -{1\over R^2}\right){\tilde \mu}^{(1)}_1=-
{hr_0^2\over ML} \Phi(R)  \left( {\tilde v}_{r1}^{(1)}(r)-{U\over 2} \right)\ .
\label{eqn:Drhoeq}\end{equation}
Equation (\ref{eqn:musurface}) gives ${\bm e}_r\cdot\nabla \mu^{(1)}=0$ at $r=r_0$, and
${\tilde \mu}^{(1)}_1(r)$ tends to zero as $r\to\infty$.  
Using Eq.~(\ref{eqn:vr1}), we can rewrite the rhs of Eq.~(\ref{eqn:Drhoeq}) as
\begin{equation}
-{h r_0^2 U\over 2\chi ML} \Phi(R)  \left( \left[{\cal L} {\hat A}\right](R) 
-\chi \right)\ .
\label{eqn:Omega}\end{equation}
Regarding Eq.~(\ref{eqn:Drhoeq}) as an inhomogeneous linear equation with constant
coefficients, we can solve it formally. 
Writing $\Omega(R)$ for Eq.~(\ref{eqn:Omega}) here, we have
\begin{equation}
{\tilde \mu}_1^{(1)}(r)=\left[{\Delta}_1^{-1}\Omega \right]\hspace{-0.5mm}(R) 
\ .\label{eqn:tilmusol}
\end{equation}  As shown by Eq.~(\ref{eqn:delinvdef}), 
$\Delta_1^{-1}$ involves the kernel $G$, which is the same as given by Eq.~(3.57) of \citet{oldraft}.
Substituting Eq.~(\ref{eqn:tilmusol}) into the rhs of Eq.~(\ref{eqn:calAout}) yields Eq.~(\ref{eqn:calAout2}).  
When $\kappa$ vanishes, Eq.~(\ref{eqn:calAin}) and (\ref{eqn:calAout})
are reduced to Eq.~(3.26) of \citet{oldraft}, whose
${\cal A}(\zeta)$ and $\kappa(\rho)$ should be respectively read as $A(\zeta)/(2\eta_3 r_0^2 U)$ and $\Phi(R)$ defined
in the present study.\\

As mentioned in Sect.~\ref{sec:critical},  
we in general have Eq.~(\ref{eqn:hatmudef}) with $\mu^{(0)}$ being replaced by $\mu({\bm r})$, and thus
$\left( m- M\Delta \right) \varphi^{(1)}=\mu^{(1)}$.
Equation (\ref{eqn:phisurface}) yields ${\bm e}_r\cdot\nabla\varphi^{(1)}=0$ at $r=r_0$, while 
$\varphi^{(1)}(r)$ tends to zero as $r\to\infty$.  
We thus have
\begin{equation}
{\tilde \varphi}_1^{(1)}(r)=-{r_0^2\over M}
\int_1^\infty dR'\ 
{\Gamma(R, R')\over R}{\tilde \mu}_1^{(1)}(r_0 R')
\ ,\label{eqn:phitilsol}\end{equation}
where the kernel $\Gamma$ is defined by Eq.~(3.55) of \citet{oldraft}.  
We need not know its full expression, and here use only
\begin{equation}
\Gamma(1,R)= {Rs_{\rm c} \Phi(R) K_1(s_{\rm c}^{-1}) \over K_1'(s_{\rm c}^{-1})}
\ .\end{equation}
The $x$-component of the drag force is given by the sum of Eqs.~(3.29) and (3.30) of
\citet{oldraft}.  Part of this sum is rewritten with the aid of  its Eq.~(3.31)
and the statement below its Eq.~(3.32).  Thus,
$\gamma$ is found to be given by the sum of Eq.~(\ref{eqn:simplegam}) and
\begin{equation}
{2 \pi h {\tilde \varphi}^{(1)}_1(1) \over U} \left(
1+{K_0(s_{\rm c}^{-1})\over
s_{\rm c}K_1(s_{\rm c}^{-1})}\right)
\ .\label{eqn:drag}\end{equation} 
Using Eqs.~(\ref{eqn:tilmusol}) and (\ref{eqn:phitilsol}), we can transform
Eq.~(\ref{eqn:drag}) into Eq.~(\ref{eqn:dragterm2}).  In this transformation,
it is helpful to rewrite $s_{\rm c}K_1(s_{\rm c}^{-1})$ and $K_1'(s_{\rm c}^{-1})$
in terms of $K_0(s_{\rm c}^{-1})$ and $K_2(s_{\rm c}^{-1})$. \\

For description of the recurrence relations, 
we rewrite double integrals contained in
each of $\left[ {\cal L}{\cal H}\Omega\right](R)$ and  $\left[ {\cal N}{\cal H}\Omega\right](R)$, where
$\Omega$ is a function.  In each, when the integrand converges absolutely, 
the function is continuous at $R=1$ and the order of the integrals can be 
exchanged to yield
\begin{equation}
 \left[ {\cal L}{\cal H}\Omega\right](1)=\int_1^\infty d R\ R\Phi(R)\Omega(R)\left[{\cal L}{\hat A}_0^{(0)}\right](R)
\label{eqn:LHp1}\end{equation}
and
\begin{equation}
 \left[ {\cal N}{\cal H}\Omega\right](1)=\int_1^\infty d R\ \Phi(R)\Omega(R)w(R)
\ ,\label{eqn:NHp1}\end{equation}
where
\begin{equation}
w(R)\equiv \int_0^\infty d\zeta\ {J_2'(\zeta)J_1(\zeta R)\over 1+\nu_{\rm o}\zeta}
\ .\label{eqn:wdef}\end{equation}
We also introduce \begin{equation}
T_n^{(k)}(R)\equiv  \Phi(R) \left[ \Delta_1^{-1} \Phi\left( {\cal L}{\hat A}_n^{(k)} -\delta_{n0}\delta_{k0} \chi \right) \right](R)  
\ .\label{eqn:teendef}\end{equation}
We substitute Eqs.~(\ref{tilan})-(\ref{eqn:explam}) into  
Eqs.~(\ref{eqn:Atilexpress0}), (\ref{eqn:vrU220}), and (\ref{eqn:tangcond20}) with 
Eqs.~(\ref{eqn:add}), (\ref{eqn:add2}), and (\ref{eqn:add3}) being supplemented respectively.
The results for $k=1,2,\ldots$ are as follows.  
We have $\beta_0^{(k)}=0$ and
\begin{equation}
\beta_n^{(k)}= \nu_{\rm o}
\left(\alpha_{n-1}^{(k)}\left[{\cal N}{\hat A}_0^{(0)}\right](1)+\beta_{n-1}^{(k)}g+
\left[{\cal N}{\cal M}{\hat A}_{n-2}^{(k)}\right](1)+
\int_1^\infty dR\ T_{n-1}^{(k-1)}(R)w(R)\right)\label{beta2}
\end{equation}
for $n=1,2,\ldots$, where we stipulate ${\hat A}^{(k)}_{-1}=0$. 
For $n=0,1,\ldots$, we have 
 \begin{eqnarray}
&&\alpha_n^{(k)}
=-{ 1\over \chi}\left\{ \beta_{n}^{(k)}\left[{\cal L}J_0\right](1) +
\left[ {\cal L}{\cal M}{\hat A}^{(k)}_{n-1}\right](1)+
 \int_1^\infty dR\ RT_n^{(k-1)}(R) \left[ {\cal L} {\hat A}_0^{(0)}\right] (R)\right\}
\label{alpha2} \\&&{\rm and}\quad
{\hat A}_n^{(k)}= \alpha_n^{(k)}{\hat A}_0^{(0)}
+\beta_n^{(k)}J_0+{\cal M}{\hat A}_{n-1}^{(k)} +
\left[{\cal S}T_n^{(k-1)}\right](\zeta)\ ,   \label{hata2}
\end{eqnarray}
where the operator ${\cal S}$ is so defined that
 \begin{equation}
\left[{\cal S}T_n^{(k-1)}\right](\zeta)\equiv{1\over \zeta}\int_1^\infty dR\ J_1(\zeta R) T_n^{(k-1)}(R)
\end{equation} 
holds.  For convenience of numerical calculations, we utilize Eq.~(\ref{hata2})
in Eq.~(\ref{eqn:gam}) to obtain
\begin{equation}
{\hat \gamma}_n^{(k)}=
\alpha_n^{(k)} + 2\beta_n^{(k)} +\Theta {\cal M} {\hat A}^{(k)}_{n - 1}+\int_1^\infty dR\ RT^{(k-1)}_n(R)
\label{gamman1}\end{equation}
for $n=0,1,\ldots$.  Here, we note
that $\Theta {\cal S} T_n^{(k-1)}$ vanishes because of the statement at the end of Appendix C of \citet{oldraft}.  
\\

The drag coefficient for $\kappa=0$ up to the order of $\lambda^2$ was studied
in \citet{oldraft}, whose Eq.~(3.25) should have had 
${\bm v}^{(1)}-U{\bm e}_x$ instead of ${\bm v}^{(1)}$ \citep{erratum}.  
The corrected equation generates Eq.~(\ref{eqn:Drhoeq}) of the present study.  
The missed term, $-U{\bm e}_x$, is found 
to generate the term involving $\chi$ in Eq.~(\ref{eqn:teendef}), 
which is contained in the first and fourth terms on the rhs of Eq.~(\ref{gamman1}) for $n=0$ and $k=1$.
In particular, $\chi$ in the fourth term makes ${\hat\gamma}_0^{(1)}$ positive;
this positivity is shown in Fig.~\ref{fig:tani}(a).  
The factor $d$ in Eq.~(3.59) of \citet{oldraft} was erroneously found to be negative because of the missed term.
This factor should equal ${\hat\gamma}_0^{(1)}$ in the present study; it
is positive and is roughly proportional to $s_{\rm c}^4$ for small $s_{\rm c}$, as shown in Fig.~\ref{fig:kdep}(a).

\section{Numerical procedure\label{app:num}}
In the recurrence relations,  
we encounter various integrands each of which contains a function generated after the operation of  
${\cal M}$.  For example, ${\cal N}{\cal M}{\hat A}_{n-2}^{(0)}$
and ${\cal L}{\cal M}{\hat A}_{n-1}^{(0)}$ in Eqs.~(\ref{beta}) and (\ref{alpha})
contain this kind of integrands.  
It is helpful in calculating the integration over a semi-infinite interval numerically
to find how the integrand behaves for its large variable.
As described in Appendix C of \citet{tani}, part of
${\cal M}J_0$ has a peculiar logarithmic dependence for its large variable, although $[{\cal N}{\cal M}J_0](R)$ is continuous at $R = 1$.
Let us define ${\hat B}_n^{(0)}$ as
${\hat A}_n^{(0)}-\beta_n^{(0)} J_0$ to calculate ${\cal M}J_0$ separately.
We find that $\left[{\cal M}{\hat B}_n^{(0)}\right]\left(\xi\right)$ is asymptotically proportional to
\begin{equation}
{1\over \xi\sqrt{\xi}}\cos{\left(\xi+\delta\right)}
\label{eqn:cossqrt}\end{equation}
as $\xi$ becomes large.
We can fix the constant of proportionality and the phase shift $\delta$
from the numerical calculation of the function for relatively small $\xi$.
The phase shift is close to $\pi/4$, as suggested in \citet{almost}.  \\
   
We can replace ${\cal M} {\hat A}_m^{(0)}$ by
$\beta_m^{(0)}{\cal M}J_0+{\cal M}{\hat B}_m^{(0)}$ in Eqs.~(\ref{beta}) and (\ref{alpha}).  
The term $\Theta {\cal M} {\hat A}^{(0)}_{n-1}$ in Eq.~(\ref{gamman})
is rewritten as $\beta_{n-1}\Theta {\cal M}J_0+\Theta {\cal M}{\hat B}^{(0)}_{n-1}$.
To calculate $\Theta{\cal M}{\hat B}_{n-1}^{(0)}$ numerically,
we can utilize Eq.~(C3) of Ref,~\citet{tani}, as suggested below this equation.
It is thus convenient to use
\begin{equation}
{\cal M}{\hat B}_n^{(0)}=\alpha_n^{(0)}{\cal M}{\hat A}_0^{(0)}+\beta_{n-1}^{(0)} \left[{\cal M}{\cal M}J_0\right]
+\left[ {\cal M}{\cal M}{\hat B}_{n-1}^{(0)}\right]\ ,
\end{equation} 
instead of Eq.~(\ref{hata}), 
in the recursion relations for $\lambda=0$.  
The last term above also asymptotically proportional to
Eq.~(\ref{eqn:cossqrt}) for large variable. \\

The above procedure is also available in the recursion relations for $\lambda\ne 0$, {\it i.e.\/},
Eqs.~(\ref{beta2})-(\ref{hata2}) and (\ref{gamman1}).  
The term ${\cal L}{\hat A}_n^{(0)}$ in Eq.~(\ref{eqn:teendef}) for $k=0$ is 
calculated by means of Eq.~(\ref{hata}), whose last term can be rewritten as
$\beta_{n-1}^{(0)} \left[{\cal M}J_0\right]+\left[{\cal M}{\hat B}_{n-1}^{(0)}\right]$. 
As $\xi$ increases, $\left[{\cal M}{\cal S}T^{(1)}_n\right](\xi)$ is asymptotically proportional to
Eq.~(\ref{eqn:cossqrt}) with
$\delta$ being close to $3\pi/4$. 

\section{Transport coefficients \label{app:mct}}
By considering the equilibrium fluctuation of a near-critical fluid, 
one finds that the transport coefficient 
on large length scales should be affected
by the convection due to long-lived correlated clusters smaller than the correlation length \citep{Onukibook}. 
The Onsager coefficient $L$ appearing in Eq.~(\ref{eqn:phidyn})
should be regarded as already coarse-grained up to the correlation length, denoted by $\xi_{\rm c}$.  
We can find how $L$ depends on $\xi_{\rm c}$ by applying the mode-coupling theory,
where the nonlinear dynamics is projected on the linear dynamics with the
nonlinear terms being considered up to the second order \citep{Onukibook, mori, Kawa, zwanzig}.
The calculation procedure used in the application of the mode-coupling theory
for the model H is here available, 
except that the Oseen tensor should be modified to fit the fluid membrane immersed in a 3D fluid, as
discussed in \citet{hydro}.  The experimental results in \citet{prl} is well explained by
the theoretical results derived in \citet{inafuji}, which
are below improved in some points.  \\ 

The fluid membrane considered here is the same as
considered in Sect.~\ref{sec:critical}, except that it has no circular liquid domain
and that it has the critical composition.  The order parameter $\psi$ is
defined as the deviation of $\varphi$ from its critical value.
From the first term of Eq.~(\ref{eqn:glw}),  the free-energy functional corresponding with the grand potential is found to be
the integral of $\left(m\psi^2+M\left|\nabla\psi\right|^2\right)/2$ over the membrane.
We assume the dependence of $\psi$ on ${\bm r}$ and the time $t$
to study its local fluctuation, write $C({\bm r}, t)$ for 
the equilibrium average of $\psi({\bm r}, t)\psi({\bm 0},0)$,
and below consider its Fourier transform  
 with respect to $x$ and $y$-components of ${\bm r}$.  The  
relaxation constant of the Fourier transform with the 2D wavenumber vector ${\bm q}$
is given by the sum of the van Hove term involving the bare value of $L$, denoted by $L^{(b)}$, and
\begin{equation}
{k_{\rm B} T\over 4\pi^2 } \int d{\bm p}\ {p^2q^2-\left({\bm p}\cdot{\bm q}\right)^2
\over 2\eta_3  \left| {\bm p}-{\bm q} \right|+\eta_{\rm o} \left| {\bm p}-{\bm q} \right|^2}
{1+q^2\xi_{\rm c}^2\over 1+p^2\xi_{\rm c}^2}
\ ,\label{eqn:inafuji}\end{equation}
where $p$ and $q$ respectively denote $\left| {\bm p} \right|$ and $\left| {\bm q} \right|$.
The above is equivalent to Eq.~(2.21) of \citet{inafuji}; the infra-red cutoff of the integration is put equal to be
zero, as in the derivation of the Kawasaki function in the mode-coupling theory for the model H. 
The van Hove term is given by $M L^{(b)} q^2\left(q^2+\xi_{\rm c}^{-2}\right)$, as mentioned in
\citet{inafuji}.  The Onsager coefficient 
$L$ appearing in Eq.~(\ref{eqn:phidyn}) is so defined that 
the sum of the two terms are equal to the van Hove term with $L^{(b)}$ being replaced by $L$.  Thus,
the quotient of Eq.~(\ref{eqn:inafuji}) divided by $M q^2 \left(q^2+\xi_{\rm c}^{-2}\right)$ equals $L-L^{(b)}$,
where the microscopic value $L^{(b)}$ is expected to be much smaller than $L$.
The interdiffusion coefficient, denoted by $D$, is defined as the quotient of the relaxation
constant divided by $q^2$.  \\

In Eq.~(\ref{eqn:inafuji}), we change the variable of the integration into ${\bm K}\equiv \left({\bm q}-{\bm p}\right)\xi_{\rm c}$,
and  the integration with respect to its angular component of ${\bm K}$ gives 
\begin{equation}
D={k_{\rm B}T\left(1+Q^{-2}\right) \over 8\pi\eta_{\rm o}} \int_0^\infty dK\ 
{ Q^2+K^2+1-\sqrt{\left( Q^2+K^2+1\right)^2-4Q^2K^2}\over K^2\left( 
K+N \right)}\ ,\label{eqn:newinafuji}
 \end{equation} where
we use $Q\equiv q\xi_{\rm c}$ and $N\equiv 2\eta_3\xi_{\rm c}/\eta_{\rm o}=s_{\rm c}/\nu_{\rm o}$.
When $Q$ is much smaller than unity, $D$ is approximately equal to
$ML/\xi_{\rm c}^2$, while Eq.~(\ref{eqn:newinafuji}) leads to
\begin{equation}
D\approx {k_{\rm B}T\over 4\pi \eta_{\rm o}} \int_0^\infty dK \ {1\over \left(K+N\right)\left(K^2+1\right)}
={k_{\rm B}T\over 4\pi \eta_{\rm o}} {\pi N-2\ln{N}\over 2\left(1+N^2\right)} 
\ ,\label{eqn:inafuji0}\end{equation}
which is equivalent to Eq.~(3.2) of \citet{inafuji}.  
When $Q$ is much larger than unity, Eq.~(\ref{eqn:newinafuji}) leads to
\begin{eqnarray}&&
D\approx {k_{\rm B}T\over 4\pi \eta_{\rm o}} \left[
\int_0^Q dK \ {1\over K+N}+\int_Q^\infty dK\ {Q^2\over K^2\left(K+N\right)} \right]\\
&& = {k_{\rm B}T\over 4\pi \eta_{\rm o}} 
\left[ \ln{\left(1+{Q\over N}\right)} + {Q\over N} -{Q^2\over N^2} \ln{\left(1+{N\over Q}\right)}\right]
\ ,\label{eqn:highapp}\end{eqnarray}
which is newly derived here.
Figure \ref{fig:inaf} shows that these approximations work well. 
For $N\ll Q$, the sum in the braces on the rhs of Eq.~(\ref{eqn:highapp}) is approximately equal to 
$\left(1/2\right)+\ln{\left(Q/N\right)}$.  
\\

\begin{figure}[t]
\begin{center}
\includegraphics[width=6cm]{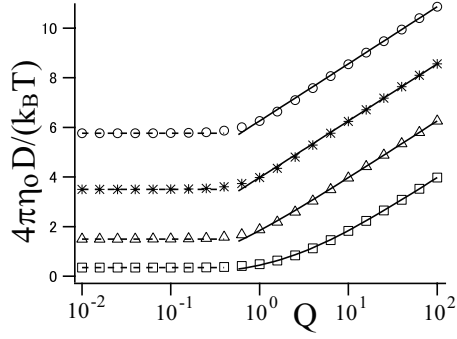}
\end{center}
\caption{Circles, asterisks, triangles, and squares represent Eq.~(\ref{eqn:newinafuji}) multiplied by
$4\pi\eta_{\rm o}/(k_{\rm B}T)$ for $N=3.2\times 10^{-3}$, $\times10^{-2}$, $\times 10^{-1}$, and
$\times 10^0$, respectively.  The solid curves from the top to the bottom represent the sum in the braces on the rhs of
 Eq.~(\ref{eqn:highapp}) for the values of $N$ above, respectively.  The dashed lines from the top to the
bottom represent the second fraction on the rhs of Eq.~(\ref{eqn:inafuji0}) for the values of $N$, respectively. }
\label{fig:inaf}
\end{figure}

We below
calculate the dependence of $\eta_{\rm o}$ on $\xi_{\rm c}$. 
Writing $\eta_{\rm o}^{(b)}$ for its bare value, we apply the mode-coupling theory to
find that $\eta_{\rm o}-\eta_{\rm o}^{(b)}$ is given by \citep{Onukibook}
\begin{equation}
{M^2\over k_{\rm B}T} \int_0^\infty dt\int d{\bm r} \ 
\left[{\partial^2  C\over \partial x^2}{\partial^2  C\over \partial y^2}
+\left( {\partial^2 C\over \partial x\partial y}\right)^2\right]
\ ,\end{equation}
where $C$ implies $C({\bm r}, t)$ and the integral with respect to ${\bm r}$
should be taken over the length scales from the microscopic
scale (denoted by $l_0$) to $\xi_{\rm c}$.  The above can be rewritten as
\begin{equation}
{k_{\rm B}T \over 8\pi} \int_{1/\xi_{\rm c}}^{1/l_0} dq {q^5 \over \left(q^2+\xi_{\rm c}^{-2}\right)^2} \int_0^\infty
dt\ e^{-2Dq^2 t}
\ .\label{eqn:etar}\end{equation}
When $N$ is much smaller than unity, {\it i.e.},  when $\xi_{\rm c}$ is much smaller than $\eta_{\rm o}/(2\eta_3)$,
we use the approximation for $D$ mentioned at the end of
the preceding paragraph to obtain
\begin{eqnarray}
&&\eta_{\rm o} \approx \eta_{\rm o}^{(b)}\left[ 1+\int_1^{\xi_{\rm c}/l_0} dQ\ {1\over 2Q+4Q\ln{\left(Q/N\right)} }\right]
\label{eqn:etar2a}\\
&& \qquad \propto \left[ 1+ {\ln{\left(\xi_{\rm c}/l_0\right)}\over (1/2)-\ln{N} }\right]^{1/4}
\approx  \left[  \ln{ {\eta_{\rm o}^{(b)}\over 2\eta_3\xi_{\rm c}} }\right]^{-1/4}
\ ,\label{eqn:etar2}\end{eqnarray}
which represents weak dependence of $\eta_{\rm o}$ on $\xi_{\rm c}$.
We can calculate the integral of Eq.~(\ref{eqn:etar2a}) numerically by using the rhs of Eq.~(\ref{eqn:highapp})
even when $N\ll Q$ cannot be assumed in the integrand.
According to the results not shown here, $\eta_{\rm o}$ triples at the most
as $\xi_{\rm c}$ changes from $1$ to $100\ $nm for practically possible material constants.
Thus, for the parameter values considered in the text, 
the dependence of $\eta$ on $\xi_{\rm c}$ can be neglected.  
In passing, the corresponding calculation in the model H for a 3D mixture
gives the shear viscosity proportional to the correlation length to the power
$8/(15\pi^2)$ \citep{Onukibook,ohta,ohtaKawa}.  The power is given by $(d-2)/3$ for an isolated $d$-dimensional fluid,
when $d$ is the real-valued dimension close to and larger than two \citep{ohta, ohtaKawa}.  

\bibliography{critraftdrag2-references}

\providecommand{\noopsort}[1]{}\providecommand{\singleletter}[1]{#1}%
\begin{thebibliography}{68}%
\makeatletter
\providecommand \@ifxundefined [1]{%
 \@ifx{#1\undefined}
}%
\providecommand \@ifnum [1]{%
 \ifnum #1\expandafter \@firstoftwo
 \else \expandafter \@secondoftwo
 \fi
}%
\providecommand \@ifx [1]{%
 \ifx #1\expandafter \@firstoftwo
 \else \expandafter \@secondoftwo
 \fi
}%
\providecommand \natexlab [1]{#1}%
\providecommand \enquote  [1]{``#1''}%
\providecommand \bibnamefont  [1]{#1}%
\providecommand \bibfnamefont [1]{#1}%
\providecommand \citenamefont [1]{#1}%
\providecommand \href@noop [0]{\@secondoftwo}%
\providecommand \href [0]{\begingroup \@sanitize@url \@href}%
\providecommand \@href[1]{\@@startlink{#1}\@@href}%
\providecommand \@@href[1]{\endgroup#1\@@endlink}%
\providecommand \@sanitize@url [0]{\catcode `\\12\catcode `\$12\catcode
  `\&12\catcode `\#12\catcode `\^12\catcode `\_12\catcode `\%12\relax}%
\providecommand \@@startlink[1]{}%
\providecommand \@@endlink[0]{}%
\providecommand \url  [0]{\begingroup\@sanitize@url \@url }%
\providecommand \@url [1]{\endgroup\@href {#1}{\urlprefix }}%
\providecommand \urlprefix  [0]{URL }%
\providecommand \Eprint [0]{\href }%
\providecommand \doibase [0]{http://dx.doi.org/}%
\providecommand \selectlanguage [0]{\@gobble}%
\providecommand \bibinfo  [0]{\@secondoftwo}%
\providecommand \bibfield  [0]{\@secondoftwo}%
\providecommand \translation [1]{[#1]}%
\providecommand \BibitemOpen [0]{}%
\providecommand \bibitemStop [0]{}%
\providecommand \bibitemNoStop [0]{.\EOS\space}%
\providecommand \EOS [0]{\spacefactor3000\relax}%
\providecommand \BibitemShut  [1]{\csname bibitem#1\endcsname}%
\let\auto@bib@innerbib\@empty
\bibitem [{\citenamefont {Sutherland}(1905)}]{suther}%
  \BibitemOpen
  \bibfield  {author} {\bibinfo {author} {\bibfnamefont {W.}~\bibnamefont
  {Sutherland}},\ }\href@noop {} {\bibfield  {journal} {\bibinfo  {journal}
  {Phil.~Mag.~}\ }\textbf {\bibinfo {volume} {9}},\ \bibinfo {pages} {781}
  (\bibinfo {year} {1905})}\BibitemShut {NoStop}%
\bibitem [{\citenamefont {Einstein}(1905)}]{eins}%
  \BibitemOpen
  \bibfield  {author} {\bibinfo {author} {\bibfnamefont {A.}~\bibnamefont
  {Einstein}},\ }\href@noop {} {\bibfield  {journal} {\bibinfo  {journal}
  {Ann.~Phys.~(Leipzig)}\ }\textbf {\bibinfo {volume} {322}},\ \bibinfo {pages}
  {549} (\bibinfo {year} {1905})}\BibitemShut {NoStop}%
\bibitem [{\citenamefont {Bian}\ \emph {et~al.}(2016)\citenamefont {Bian},
  \citenamefont {Kim},\ and\ \citenamefont {Karniadakis}}]{111years}%
  \BibitemOpen
  \bibfield  {author} {\bibinfo {author} {\bibfnamefont {X.}~\bibnamefont
  {Bian}}, \bibinfo {author} {\bibfnamefont {C.}~\bibnamefont {Kim}}, \ and\
  \bibinfo {author} {\bibfnamefont {G.~E.}\ \bibnamefont {Karniadakis}},\
  }\href@noop {} {\bibfield  {journal} {\bibinfo  {journal} {Soft Matter}\
  }\textbf {\bibinfo {volume} {12}},\ \bibinfo {pages} {6331} (\bibinfo {year}
  {2016})}\BibitemShut {NoStop}%
\bibitem [{\citenamefont {Waigh}(2005)}]{MicroRev}%
  \BibitemOpen
  \bibfield  {author} {\bibinfo {author} {\bibfnamefont {T.~A.}\ \bibnamefont
  {Waigh}},\ }\href@noop {} {\bibfield  {journal} {\bibinfo  {journal}
  {Rep.~Prog.~Phys.~}\ }\textbf {\bibinfo {volume} {68}},\ \bibinfo {pages}
  {685} (\bibinfo {year} {2005})}\BibitemShut {NoStop}%
\bibitem [{\citenamefont {Dom{\'{\i}}nguez-Garc{\'{\i}}a}\ \emph
  {et~al.}(2014)\citenamefont {Dom{\'{\i}}nguez-Garc{\'{\i}}a}, \citenamefont
  {Cardinaux}, \citenamefont {Bertseva}, \citenamefont {Forr{\'o}},
  \citenamefont {Scheffold},\ and\ \citenamefont {Jeney}}]{viscel2}%
  \BibitemOpen
  \bibfield  {author} {\bibinfo {author} {\bibfnamefont {P.}~\bibnamefont
  {Dom{\'{\i}}nguez-Garc{\'{\i}}a}}, \bibinfo {author} {\bibfnamefont
  {F.}~\bibnamefont {Cardinaux}}, \bibinfo {author} {\bibfnamefont
  {E.}~\bibnamefont {Bertseva}}, \bibinfo {author} {\bibfnamefont
  {L.}~\bibnamefont {Forr{\'o}}}, \bibinfo {author} {\bibfnamefont
  {F.}~\bibnamefont {Scheffold}}, \ and\ \bibinfo {author} {\bibfnamefont
  {S.}~\bibnamefont {Jeney}},\ }\href@noop {} {\bibfield  {journal} {\bibinfo
  {journal} {Phys.~Rev.~E}\ }\textbf {\bibinfo {volume} {90}},\ \bibinfo
  {pages} {060301} (\bibinfo {year} {2014})}\BibitemShut {NoStop}%
\bibitem [{\citenamefont {Glazier}\ and\ \citenamefont {Salaita}(2017)}]{glaz}%
  \BibitemOpen
  \bibfield  {author} {\bibinfo {author} {\bibfnamefont {R.}~\bibnamefont
  {Glazier}}\ and\ \bibinfo {author} {\bibfnamefont {K.}~\bibnamefont
  {Salaita}},\ }\href@noop {} {\bibfield  {journal} {\bibinfo  {journal}
  {Biophys.~Biochim.~Acta}\ }\textbf {\bibinfo {volume} {1859}},\ \bibinfo
  {pages} {1465} (\bibinfo {year} {2017})}\BibitemShut {NoStop}%
\bibitem [{\citenamefont {Ortega}\ \emph {et~al.}(2010)\citenamefont {Ortega},
  \citenamefont {Ritacco},\ and\ \citenamefont {Rubio}}]{ortega}%
  \BibitemOpen
  \bibfield  {author} {\bibinfo {author} {\bibfnamefont {F.}~\bibnamefont
  {Ortega}}, \bibinfo {author} {\bibfnamefont {H.}~\bibnamefont {Ritacco}}, \
  and\ \bibinfo {author} {\bibfnamefont {R.~G.}\ \bibnamefont {Rubio}},\
  }\href@noop {} {\bibfield  {journal} {\bibinfo  {journal}
  {Curr.~Opin.~Colloid Interface Sci.~}\ }\textbf {\bibinfo {volume} {15}},\
  \bibinfo {pages} {237} (\bibinfo {year} {2010})}\BibitemShut {NoStop}%
\bibitem [{\citenamefont {Franosch}\ \emph {et~al.}(2011)\citenamefont
  {Franosch}, \citenamefont {Grimm}, \citenamefont {Belushkin}, \citenamefont
  {Mor}, \citenamefont {Foffi}, \citenamefont {Forr{\' o}},\ and\ \citenamefont
  {Jeney}}]{nat}%
  \BibitemOpen
  \bibfield  {author} {\bibinfo {author} {\bibfnamefont {T.}~\bibnamefont
  {Franosch}}, \bibinfo {author} {\bibfnamefont {M.}~\bibnamefont {Grimm}},
  \bibinfo {author} {\bibfnamefont {M.}~\bibnamefont {Belushkin}}, \bibinfo
  {author} {\bibfnamefont {F.~M.}\ \bibnamefont {Mor}}, \bibinfo {author}
  {\bibfnamefont {G.}~\bibnamefont {Foffi}}, \bibinfo {author} {\bibfnamefont
  {L.}~\bibnamefont {Forr{\' o}}}, \ and\ \bibinfo {author} {\bibfnamefont
  {S.}~\bibnamefont {Jeney}},\ }\href@noop {} {\bibfield  {journal} {\bibinfo
  {journal} {Nature}\ }\textbf {\bibinfo {volume} {478}},\ \bibinfo {pages}
  {85} (\bibinfo {year} {2011})}\BibitemShut {NoStop}%
\bibitem [{\citenamefont {Grimm}\ \emph {et~al.}(2012)\citenamefont {Grimm},
  \citenamefont {Franosch},\ and\ \citenamefont {Jeney}}]{grim}%
  \BibitemOpen
  \bibfield  {author} {\bibinfo {author} {\bibfnamefont {M.}~\bibnamefont
  {Grimm}}, \bibinfo {author} {\bibfnamefont {T.}~\bibnamefont {Franosch}}, \
  and\ \bibinfo {author} {\bibfnamefont {S.}~\bibnamefont {Jeney}},\
  }\href@noop {} {\bibfield  {journal} {\bibinfo  {journal} {Phys.~Rev.~E}\
  }\textbf {\bibinfo {volume} {86}},\ \bibinfo {pages} {021912} (\bibinfo
  {year} {2012})}\BibitemShut {NoStop}%
\bibitem [{\citenamefont {Singer}\ and\ \citenamefont
  {Nicolson}(1972)}]{singer}%
  \BibitemOpen
  \bibfield  {author} {\bibinfo {author} {\bibfnamefont {S.~J.}\ \bibnamefont
  {Singer}}\ and\ \bibinfo {author} {\bibfnamefont {G.~L.}\ \bibnamefont
  {Nicolson}},\ }\href@noop {} {\bibfield  {journal} {\bibinfo  {journal}
  {Science}\ }\textbf {\bibinfo {volume} {175}},\ \bibinfo {pages} {720}
  (\bibinfo {year} {1972})}\BibitemShut {NoStop}%
\bibitem [{\citenamefont {Komura}\ and\ \citenamefont
  {Andelman}(2014)}]{andel}%
  \BibitemOpen
  \bibfield  {author} {\bibinfo {author} {\bibfnamefont {S.}~\bibnamefont
  {Komura}}\ and\ \bibinfo {author} {\bibfnamefont {D.}~\bibnamefont
  {Andelman}},\ }\href@noop {} {\bibfield  {journal} {\bibinfo  {journal}
  {Adv.~Colloid Interface Sci.~}\ }\textbf {\bibinfo {volume} {208}},\ \bibinfo
  {pages} {34} (\bibinfo {year} {2014})}\BibitemShut {NoStop}%
\bibitem [{\citenamefont {Saffman}\ and\ \citenamefont {Delbr{\"
  u}ck}(1975)}]{saffman}%
  \BibitemOpen
  \bibfield  {author} {\bibinfo {author} {\bibfnamefont {P.~G.}\ \bibnamefont
  {Saffman}}\ and\ \bibinfo {author} {\bibfnamefont {M.}~\bibnamefont {Delbr{\"
  u}ck}},\ }\href@noop {} {\bibfield  {journal} {\bibinfo  {journal}
  {Proc.~Natl.~Acad.~Sci.~U. S. A.}\ }\textbf {\bibinfo {volume} {72}},\
  \bibinfo {pages} {3111} (\bibinfo {year} {1975})}\BibitemShut {NoStop}%
\bibitem [{\citenamefont {Saffman}(1976)}]{saffman2}%
  \BibitemOpen
  \bibfield  {author} {\bibinfo {author} {\bibfnamefont {P.~G.}\ \bibnamefont
  {Saffman}},\ }\href@noop {} {\bibfield  {journal} {\bibinfo  {journal} {J.
  Fluid Mech.~}\ }\textbf {\bibinfo {volume} {73}},\ \bibinfo {pages} {593}
  (\bibinfo {year} {1976})}\BibitemShut {NoStop}%
\bibitem [{\citenamefont {Hughes}\ \emph {et~al.}(1981)\citenamefont {Hughes},
  \citenamefont {Pailthorpe},\ and\ \citenamefont {White}}]{hughes}%
  \BibitemOpen
  \bibfield  {author} {\bibinfo {author} {\bibfnamefont {B.~D.}\ \bibnamefont
  {Hughes}}, \bibinfo {author} {\bibfnamefont {B.~A.}\ \bibnamefont
  {Pailthorpe}}, \ and\ \bibinfo {author} {\bibfnamefont {L.~R.}\ \bibnamefont
  {White}},\ }\href@noop {} {\bibfield  {journal} {\bibinfo  {journal} {J.
  Fluid Mech.~}\ }\textbf {\bibinfo {volume} {110}},\ \bibinfo {pages} {349}
  (\bibinfo {year} {1981})}\BibitemShut {NoStop}%
\bibitem [{\citenamefont {Simons}\ and\ \citenamefont {Ikonen}(1997)}]{ikonen}%
  \BibitemOpen
  \bibfield  {author} {\bibinfo {author} {\bibfnamefont {K.}~\bibnamefont
  {Simons}}\ and\ \bibinfo {author} {\bibfnamefont {E.}~\bibnamefont
  {Ikonen}},\ }\href@noop {} {\bibfield  {journal} {\bibinfo  {journal}
  {Nature}\ }\textbf {\bibinfo {volume} {387}},\ \bibinfo {pages} {569}
  (\bibinfo {year} {1997})}\BibitemShut {NoStop}%
\bibitem [{\citenamefont {Leslie}(2011)}]{leslie}%
  \BibitemOpen
  \bibfield  {author} {\bibinfo {author} {\bibfnamefont {M.}~\bibnamefont
  {Leslie}},\ }\href@noop {} {\bibfield  {journal} {\bibinfo  {journal}
  {Science}\ }\textbf {\bibinfo {volume} {334}},\ \bibinfo {pages} {1046}
  (\bibinfo {year} {2011})}\BibitemShut {NoStop}%
\bibitem [{\citenamefont {Dietrich}\ \emph {et~al.}(2001)\citenamefont
  {Dietrich}, \citenamefont {Bagatolli}, \citenamefont {Volovyk}, \citenamefont
  {Thompson}, \citenamefont {M.~Levi},\ and\ \citenamefont
  {Gratton}}]{Dietrich}%
  \BibitemOpen
  \bibfield  {author} {\bibinfo {author} {\bibfnamefont {C.}~\bibnamefont
  {Dietrich}}, \bibinfo {author} {\bibfnamefont {L.~A.}\ \bibnamefont
  {Bagatolli}}, \bibinfo {author} {\bibfnamefont {Z.~N.}\ \bibnamefont
  {Volovyk}}, \bibinfo {author} {\bibfnamefont {N.~L.}\ \bibnamefont
  {Thompson}}, \bibinfo {author} {\bibfnamefont {K.~J.}\ \bibnamefont
  {M.~Levi}}, \ and\ \bibinfo {author} {\bibfnamefont {E.}~\bibnamefont
  {Gratton}},\ }\href@noop {} {\bibfield  {journal} {\bibinfo  {journal}
  {Biophys. J.}\ }\textbf {\bibinfo {volume} {80}},\ \bibinfo {pages} {1417}
  (\bibinfo {year} {2001})}\BibitemShut {NoStop}%
\bibitem [{\citenamefont {Veatch}\ and\ \citenamefont
  {Keller}(2002)}]{Veatch2002}%
  \BibitemOpen
  \bibfield  {author} {\bibinfo {author} {\bibfnamefont {S.~L.}\ \bibnamefont
  {Veatch}}\ and\ \bibinfo {author} {\bibfnamefont {S.~L.}\ \bibnamefont
  {Keller}},\ }\href@noop {} {\bibfield  {journal} {\bibinfo  {journal} {Phys.
  Rev. Lett.~}\ }\textbf {\bibinfo {volume} {89}},\ \bibinfo {pages} {268101}
  (\bibinfo {year} {2002})}\BibitemShut {NoStop}%
\bibitem [{\citenamefont {Veatch}\ and\ \citenamefont
  {Keller}(2003)}]{VeatchBJ}%
  \BibitemOpen
  \bibfield  {author} {\bibinfo {author} {\bibfnamefont {S.~L.}\ \bibnamefont
  {Veatch}}\ and\ \bibinfo {author} {\bibfnamefont {S.~L.}\ \bibnamefont
  {Keller}},\ }\href@noop {} {\bibfield  {journal} {\bibinfo  {journal}
  {Biophys J.}\ }\textbf {\bibinfo {volume} {85}},\ \bibinfo {pages} {4428}
  (\bibinfo {year} {2003})}\BibitemShut {NoStop}%
\bibitem [{\citenamefont {Veach}\ and\ \citenamefont {Keller}(2005)}]{keller}%
  \BibitemOpen
  \bibfield  {author} {\bibinfo {author} {\bibfnamefont {S.~L.}\ \bibnamefont
  {Veach}}\ and\ \bibinfo {author} {\bibfnamefont {S.~L.}\ \bibnamefont
  {Keller}},\ }\href@noop {} {\bibfield  {journal} {\bibinfo  {journal}
  {Phys.~Rev.~Lett.}\ }\textbf {\bibinfo {volume} {94}},\ \bibinfo {pages}
  {148101} (\bibinfo {year} {2005})}\BibitemShut {NoStop}%
\bibitem [{\citenamefont {Veatch}\ \emph {et~al.}(2007)\citenamefont {Veatch},
  \citenamefont {Soubias}, \citenamefont {Keller},\ and\ \citenamefont
  {Gawrisch}}]{veatch1}%
  \BibitemOpen
  \bibfield  {author} {\bibinfo {author} {\bibfnamefont {S.~L.}\ \bibnamefont
  {Veatch}}, \bibinfo {author} {\bibfnamefont {O.}~\bibnamefont {Soubias}},
  \bibinfo {author} {\bibfnamefont {S.~L.}\ \bibnamefont {Keller}}, \ and\
  \bibinfo {author} {\bibfnamefont {K.}~\bibnamefont {Gawrisch}},\ }\href@noop
  {} {\bibfield  {journal} {\bibinfo  {journal} {Proc.~Natl.~Acad.~Sci.~U. S.
  A.}\ }\textbf {\bibinfo {volume} {104}},\ \bibinfo {pages} {17650} (\bibinfo
  {year} {2007})}\BibitemShut {NoStop}%
\bibitem [{\citenamefont {Koker}(1996)}]{koker}%
  \BibitemOpen
  \bibfield  {author} {\bibinfo {author} {\bibfnamefont {R.~D.}\ \bibnamefont
  {Koker}},\ }\emph {\bibinfo {title} {The Program in Biophysics}},\ \href@noop
  {} {Ph.D. thesis},\ \bibinfo  {school} {Stanford University} (\bibinfo {year}
  {1996})\BibitemShut {NoStop}%
\bibitem [{\citenamefont {Cicuta}\ \emph {et~al.}(2007)\citenamefont {Cicuta},
  \citenamefont {Keller},\ and\ \citenamefont {Veatch}}]{cicuta}%
  \BibitemOpen
  \bibfield  {author} {\bibinfo {author} {\bibfnamefont {P.}~\bibnamefont
  {Cicuta}}, \bibinfo {author} {\bibfnamefont {S.~L.}\ \bibnamefont {Keller}},
  \ and\ \bibinfo {author} {\bibfnamefont {S.~L.}\ \bibnamefont {Veatch}},\
  }\href@noop {} {\bibfield  {journal} {\bibinfo  {journal} {J. Phys. Chem.~}\
  }\textbf {\bibinfo {volume} {111}},\ \bibinfo {pages} {3328} (\bibinfo {year}
  {2007})}\BibitemShut {NoStop}%
\bibitem [{\citenamefont {Aliaskarisohi}\ \emph {et~al.}(2010)\citenamefont
  {Aliaskarisohi}, \citenamefont {Tierno}, \citenamefont {Dhar}, \citenamefont
  {Khattari}, \citenamefont {Blaszczynski},\ and\ \citenamefont
  {Fischer}}]{alias}%
  \BibitemOpen
  \bibfield  {author} {\bibinfo {author} {\bibfnamefont {S.}~\bibnamefont
  {Aliaskarisohi}}, \bibinfo {author} {\bibfnamefont {P.}~\bibnamefont
  {Tierno}}, \bibinfo {author} {\bibfnamefont {P.}~\bibnamefont {Dhar}},
  \bibinfo {author} {\bibfnamefont {Z.}~\bibnamefont {Khattari}}, \bibinfo
  {author} {\bibfnamefont {M.}~\bibnamefont {Blaszczynski}}, \ and\ \bibinfo
  {author} {\bibfnamefont {T.~M.}\ \bibnamefont {Fischer}},\ }\href@noop {}
  {\bibfield  {journal} {\bibinfo  {journal} {J. Fluid Mech.}\ }\textbf
  {\bibinfo {volume} {654}},\ \bibinfo {pages} {417} (\bibinfo {year}
  {2010})}\BibitemShut {NoStop}%
\bibitem [{\citenamefont {Fujitani}(2013{\natexlab{a}})}]{confine}%
  \BibitemOpen
  \bibfield  {author} {\bibinfo {author} {\bibfnamefont {Y.}~\bibnamefont
  {Fujitani}},\ }\href@noop {} {\bibfield  {journal} {\bibinfo  {journal} {J.
  Phys.~Soc.~Jpn.~}\ }\textbf {\bibinfo {volume} {82}},\ \bibinfo {pages}
  {084403} (\bibinfo {year} {2013}{\natexlab{a}})}\BibitemShut {NoStop}%
\bibitem [{\citenamefont {Tani}\ and\ \citenamefont {Fujitani}(2018)}]{tani}%
  \BibitemOpen
  \bibfield  {author} {\bibinfo {author} {\bibfnamefont {H.}~\bibnamefont
  {Tani}}\ and\ \bibinfo {author} {\bibfnamefont {Y.}~\bibnamefont
  {Fujitani}},\ }\href@noop {} {\bibfield  {journal} {\bibinfo  {journal} {J.
  Fluid Mech.~}\ }\textbf {\bibinfo {volume} {836}},\ \bibinfo {pages} {910}
  (\bibinfo {year} {2018})}\BibitemShut {NoStop}%
\bibitem [{\citenamefont {Honerkamp-Smith}\ \emph {et~al.}(2008)\citenamefont
  {Honerkamp-Smith}, \citenamefont {Cicuta}, \citenamefont {Collins},
  \citenamefont {Veatch}, \citenamefont {den Nijs}, \citenamefont {Schick},\
  and\ \citenamefont {Keller}}]{honerBJ}%
  \BibitemOpen
  \bibfield  {author} {\bibinfo {author} {\bibfnamefont {A.~R.}\ \bibnamefont
  {Honerkamp-Smith}}, \bibinfo {author} {\bibfnamefont {P.}~\bibnamefont
  {Cicuta}}, \bibinfo {author} {\bibfnamefont {M.~D.}\ \bibnamefont {Collins}},
  \bibinfo {author} {\bibfnamefont {S.~L.}\ \bibnamefont {Veatch}}, \bibinfo
  {author} {\bibfnamefont {M.}~\bibnamefont {den Nijs}}, \bibinfo {author}
  {\bibfnamefont {M.}~\bibnamefont {Schick}}, \ and\ \bibinfo {author}
  {\bibfnamefont {S.~L.}\ \bibnamefont {Keller}},\ }\href@noop {} {\bibfield
  {journal} {\bibinfo  {journal} {Biophys J.}\ }\textbf {\bibinfo {volume}
  {95}},\ \bibinfo {pages} {236} (\bibinfo {year} {2008})}\BibitemShut
  {NoStop}%
\bibitem [{\citenamefont {Veatch}\ \emph {et~al.}(2008)\citenamefont {Veatch},
  \citenamefont {Cicuta}, \citenamefont {Sengupta}, \citenamefont
  {Honerkamp-Smith}, \citenamefont {Holowka},\ and\ \citenamefont
  {Baird}}]{veatch2}%
  \BibitemOpen
  \bibfield  {author} {\bibinfo {author} {\bibfnamefont {S.~L.}\ \bibnamefont
  {Veatch}}, \bibinfo {author} {\bibfnamefont {P.}~\bibnamefont {Cicuta}},
  \bibinfo {author} {\bibfnamefont {P.}~\bibnamefont {Sengupta}}, \bibinfo
  {author} {\bibfnamefont {A.~R.}\ \bibnamefont {Honerkamp-Smith}}, \bibinfo
  {author} {\bibfnamefont {D.}~\bibnamefont {Holowka}}, \ and\ \bibinfo
  {author} {\bibfnamefont {B.}~\bibnamefont {Baird}},\ }\href@noop {}
  {\bibfield  {journal} {\bibinfo  {journal} {ACS Chem.~Biol.~}\ }\textbf
  {\bibinfo {volume} {3}},\ \bibinfo {pages} {287} (\bibinfo {year}
  {2008})}\BibitemShut {NoStop}%
\bibitem [{\citenamefont {Honerkamp-Smith}\ \emph {et~al.}(2009)\citenamefont
  {Honerkamp-Smith}, \citenamefont {Veatch},\ and\ \citenamefont
  {Keller}}]{honer}%
  \BibitemOpen
  \bibfield  {author} {\bibinfo {author} {\bibfnamefont {A.~R.}\ \bibnamefont
  {Honerkamp-Smith}}, \bibinfo {author} {\bibfnamefont {S.~L.}\ \bibnamefont
  {Veatch}}, \ and\ \bibinfo {author} {\bibfnamefont {S.~L.}\ \bibnamefont
  {Keller}},\ }\href@noop {} {\bibfield  {journal} {\bibinfo  {journal}
  {Biochim.~Biophys.~Acta}\ }\textbf {\bibinfo {volume} {1788}},\ \bibinfo
  {pages} {53} (\bibinfo {year} {2009})}\BibitemShut {NoStop}%
\bibitem [{\citenamefont {Hohenberg}\ and\ \citenamefont
  {Halperin}(1977)}]{Hohenberg}%
  \BibitemOpen
  \bibfield  {author} {\bibinfo {author} {\bibfnamefont {P.~C.}\ \bibnamefont
  {Hohenberg}}\ and\ \bibinfo {author} {\bibfnamefont {B.~I.}\ \bibnamefont
  {Halperin}},\ }\href@noop {} {\bibfield  {journal} {\bibinfo  {journal}
  {Rev.~Mod.~Phys.~}\ }\textbf {\bibinfo {volume} {49}},\ \bibinfo {pages}
  {435} (\bibinfo {year} {1977})}\BibitemShut {NoStop}%
\bibitem [{\citenamefont {Onuki}(2002)}]{Onukibook}%
  \BibitemOpen
  \bibfield  {author} {\bibinfo {author} {\bibfnamefont {A.}~\bibnamefont
  {Onuki}},\ }\href@noop {} {\emph {\bibinfo {title} {Phase Transition
  Dynamics}}}\ (\bibinfo  {publisher} {Cambridge University Press},\ \bibinfo
  {year} {2002})\ Chap.~\bibinfo {chapter} {6}\BibitemShut {NoStop}%
\bibitem [{\citenamefont {Inaura}\ and\ \citenamefont
  {Fujitani}(2008)}]{inafuji}%
  \BibitemOpen
  \bibfield  {author} {\bibinfo {author} {\bibfnamefont {K.}~\bibnamefont
  {Inaura}}\ and\ \bibinfo {author} {\bibfnamefont {Y.}~\bibnamefont
  {Fujitani}},\ }\href@noop {} {\bibfield  {journal} {\bibinfo  {journal} {J.
  Phys.~Soc.~Jpn.}\ }\textbf {\bibinfo {volume} {77}},\ \bibinfo {pages}
  {114603} (\bibinfo {year} {2008})}\BibitemShut {NoStop}%
\bibitem [{\citenamefont {Haataja}(2009)}]{haat}%
  \BibitemOpen
  \bibfield  {author} {\bibinfo {author} {\bibfnamefont {M.}~\bibnamefont
  {Haataja}},\ }\href@noop {} {\bibfield  {journal} {\bibinfo  {journal}
  {Phys.~Rev.~E}\ }\textbf {\bibinfo {volume} {80}},\ \bibinfo {pages} {020902}
  (\bibinfo {year} {2009})}\BibitemShut {NoStop}%
\bibitem [{\citenamefont {Honerkamp-Smith}\ \emph {et~al.}(2012)\citenamefont
  {Honerkamp-Smith}, \citenamefont {Machta},\ and\ \citenamefont
  {Keller}}]{prl}%
  \BibitemOpen
  \bibfield  {author} {\bibinfo {author} {\bibfnamefont {A.~R.}\ \bibnamefont
  {Honerkamp-Smith}}, \bibinfo {author} {\bibfnamefont {B.~B.}\ \bibnamefont
  {Machta}}, \ and\ \bibinfo {author} {\bibfnamefont {S.~L.}\ \bibnamefont
  {Keller}},\ }\href@noop {} {\bibfield  {journal} {\bibinfo  {journal}
  {Phys.~Rev.~Lett.~}\ }\textbf {\bibinfo {volume} {108}},\ \bibinfo {pages}
  {265702} (\bibinfo {year} {2012})}\BibitemShut {NoStop}%
\bibitem [{\citenamefont {Fujitani}(2013{\natexlab{b}})}]{hydro}%
  \BibitemOpen
  \bibfield  {author} {\bibinfo {author} {\bibfnamefont {Y.}~\bibnamefont
  {Fujitani}},\ }\href@noop {} {\bibfield  {journal} {\bibinfo  {journal} {J.
  Phys.~Soc.~Jpn.}\ }\textbf {\bibinfo {volume} {82}},\ \bibinfo {pages}
  {014601} (\bibinfo {year} {2013}{\natexlab{b}})}\BibitemShut {NoStop}%
\bibitem [{\citenamefont {Tserkovnyak}\ and\ \citenamefont
  {Nelson}(2006)}]{yeth}%
  \BibitemOpen
  \bibfield  {author} {\bibinfo {author} {\bibfnamefont {Y.}~\bibnamefont
  {Tserkovnyak}}\ and\ \bibinfo {author} {\bibfnamefont {D.~R.}\ \bibnamefont
  {Nelson}},\ }\href@noop {} {\bibfield  {journal} {\bibinfo  {journal}
  {Proc.~Natl.~Acad.~Sci.~U. S. A.}\ }\textbf {\bibinfo {volume} {103}},\
  \bibinfo {pages} {15002} (\bibinfo {year} {2006})}\BibitemShut {NoStop}%
\bibitem [{\citenamefont {D{\' e}mery}\ and\ \citenamefont
  {Dean}(2010)}]{demery}%
  \BibitemOpen
  \bibfield  {author} {\bibinfo {author} {\bibfnamefont {V.}~\bibnamefont {D{\'
  e}mery}}\ and\ \bibinfo {author} {\bibfnamefont {D.~S.}\ \bibnamefont
  {Dean}},\ }\href@noop {} {\bibfield  {journal} {\bibinfo  {journal}
  {Phys.~Rev.~Lett.~}\ }\textbf {\bibinfo {volume} {104}},\ \bibinfo {pages}
  {080601} (\bibinfo {year} {2010})}\BibitemShut {NoStop}%
\bibitem [{\citenamefont {Camley}\ and\ \citenamefont {Brown}(2012)}]{camley}%
  \BibitemOpen
  \bibfield  {author} {\bibinfo {author} {\bibfnamefont {B.~A.}\ \bibnamefont
  {Camley}}\ and\ \bibinfo {author} {\bibfnamefont {F.~L.~H.}\ \bibnamefont
  {Brown}},\ }\href@noop {} {\bibfield  {journal} {\bibinfo  {journal}
  {Phys.~Rev.~E}\ }\textbf {\bibinfo {volume} {85}},\ \bibinfo {pages} {061921}
  (\bibinfo {year} {2012})}\BibitemShut {NoStop}%
\bibitem [{\citenamefont {Camley}\ and\ \citenamefont {Brown}(2014)}]{camley2}%
  \BibitemOpen
  \bibfield  {author} {\bibinfo {author} {\bibfnamefont {B.~A.}\ \bibnamefont
  {Camley}}\ and\ \bibinfo {author} {\bibfnamefont {F.~L.~H.}\ \bibnamefont
  {Brown}},\ }\href@noop {} {\bibfield  {journal} {\bibinfo  {journal} {J. Chem
  Phys.~}\ }\textbf {\bibinfo {volume} {141}},\ \bibinfo {pages} {075103}
  (\bibinfo {year} {2014})}\BibitemShut {NoStop}%
\bibitem [{\citenamefont {Okamoto}\ \emph {et~al.}(2013)\citenamefont
  {Okamoto}, \citenamefont {Fujitani},\ and\ \citenamefont {Komura}}]{ofk}%
  \BibitemOpen
  \bibfield  {author} {\bibinfo {author} {\bibfnamefont {R.}~\bibnamefont
  {Okamoto}}, \bibinfo {author} {\bibfnamefont {Y.}~\bibnamefont {Fujitani}}, \
  and\ \bibinfo {author} {\bibfnamefont {S.}~\bibnamefont {Komura}},\
  }\href@noop {} {\bibfield  {journal} {\bibinfo  {journal} {J.
  Phys.~Soc.~Jpn.}\ }\textbf {\bibinfo {volume} {82}},\ \bibinfo {pages}
  {084003} (\bibinfo {year} {2013})}\BibitemShut {NoStop}%
\bibitem [{\citenamefont {Furukawa}\ \emph {et~al.}(2013)\citenamefont
  {Furukawa}, \citenamefont {Gambassi}, \citenamefont {Dietrich},\ and\
  \citenamefont {Tanaka}}]{furu}%
  \BibitemOpen
  \bibfield  {author} {\bibinfo {author} {\bibfnamefont {A.}~\bibnamefont
  {Furukawa}}, \bibinfo {author} {\bibfnamefont {A.}~\bibnamefont {Gambassi}},
  \bibinfo {author} {\bibfnamefont {S.}~\bibnamefont {Dietrich}}, \ and\
  \bibinfo {author} {\bibfnamefont {H.}~\bibnamefont {Tanaka}},\ }\href@noop {}
  {\bibfield  {journal} {\bibinfo  {journal} {Phys.~Rev.~Lett.~}\ }\textbf
  {\bibinfo {volume} {111}},\ \bibinfo {pages} {055701} (\bibinfo {year}
  {2013})}\BibitemShut {NoStop}%
\bibitem [{\citenamefont {Yabunaka}\ \emph {et~al.}(2015)\citenamefont
  {Yabunaka}, \citenamefont {Okamoto},\ and\ \citenamefont {Onuki}}]{yabu}%
  \BibitemOpen
  \bibfield  {author} {\bibinfo {author} {\bibfnamefont {S.}~\bibnamefont
  {Yabunaka}}, \bibinfo {author} {\bibfnamefont {R.}~\bibnamefont {Okamoto}}, \
  and\ \bibinfo {author} {\bibfnamefont {A.}~\bibnamefont {Onuki}},\
  }\href@noop {} {\bibfield  {journal} {\bibinfo  {journal} {Soft Matter}\
  }\textbf {\bibinfo {volume} {11}},\ \bibinfo {pages} {5738} (\bibinfo {year}
  {2015})}\BibitemShut {NoStop}%
\bibitem [{\citenamefont {Fujitani}(2013{\natexlab{c}})}]{oldraft}%
  \BibitemOpen
  \bibfield  {author} {\bibinfo {author} {\bibfnamefont {Y.}~\bibnamefont
  {Fujitani}},\ }\href@noop {} {\bibfield  {journal} {\bibinfo  {journal} {J.
  Phys.~Soc.~Jpn.}\ }\textbf {\bibinfo {volume} {82}},\ \bibinfo {pages}
  {124601} (\bibinfo {year} {2013}{\natexlab{c}})},\ \bibinfo {note} {[erratum]
  \textbf{83}, 088001 (2014)}\BibitemShut {NoStop}%
\bibitem [{pde()}]{pdel}%
  \BibitemOpen
  \href@noop {} {\ }\bibinfo {note} {We rewrite ${\tilde p}^{(1)}_1$, contained
  in Eq.~(\ref{eqn:TAU}), by using the Fourier transform of the
  $\theta$-component with the order of $\varepsilon$ extracted from the first
  equation of Eq.~(\ref{eqn:2st})}\BibitemShut {NoStop}%
\bibitem [{pre()}]{prevgam}%
  \BibitemOpen
  \href@noop {} {\ }\bibinfo {note} {Equation (\ref{eqn:simplegam}) can be
  derived from Eqs.~(2.24), (2.41) and (3.13) of \citet{confine}}\BibitemShut
  {NoStop}%
\bibitem [{com()}]{comchi}%
  \BibitemOpen
  \href@noop {} {\ }\bibinfo {note} {The integral denoted by $\chi$ is referred
  to as $\chi(1)$ in \citet{oldraft} and as $Y_0$ in \citet{tani}}\BibitemShut
  {NoStop}%
\bibitem [{gme()}]{gmethod}%
  \BibitemOpen
  \href@noop {} {\ }\bibinfo {note} {Numerical procedure of calculating $g$ is
  described in Sect.~4.2 of \citet{tani}}\BibitemShut {NoStop}%
\bibitem [{\citenamefont {Fujitani}(2014{\natexlab{a}})}]{wetdrop}%
  \BibitemOpen
  \bibfield  {author} {\bibinfo {author} {\bibfnamefont {Y.}~\bibnamefont
  {Fujitani}},\ }\href@noop {} {\bibfield  {journal} {\bibinfo  {journal}
  {J.~Phys.~Soc.~Jpn.~}\ }\textbf {\bibinfo {volume} {83}},\ \bibinfo {pages}
  {024401} (\bibinfo {year} {2014}{\natexlab{a}})},\ \bibinfo {note} {[erratum]
  \textbf{83}, 108001 (2014)}\BibitemShut {NoStop}%
\bibitem [{\citenamefont {Cahn}(1976)}]{Cahn}%
  \BibitemOpen
  \bibfield  {author} {\bibinfo {author} {\bibfnamefont {J.~W.}\ \bibnamefont
  {Cahn}},\ }\href@noop {} {\bibfield  {journal} {\bibinfo  {journal} {J. Chem.
  Phys.}\ }\textbf {\bibinfo {volume} {66}},\ \bibinfo {pages} {3667} (\bibinfo
  {year} {1976})}\BibitemShut {NoStop}%
\bibitem [{\citenamefont {Diehl}\ and\ \citenamefont {Janssen}(1992)}]{jans}%
  \BibitemOpen
  \bibfield  {author} {\bibinfo {author} {\bibfnamefont {H.~W.}\ \bibnamefont
  {Diehl}}\ and\ \bibinfo {author} {\bibfnamefont {H.~K.}\ \bibnamefont
  {Janssen}},\ }\href@noop {} {\bibfield  {journal} {\bibinfo  {journal}
  {Phys.~Rev.~A}\ }\textbf {\bibinfo {volume} {45}},\ \bibinfo {pages} {7145}
  (\bibinfo {year} {1992})}\BibitemShut {NoStop}%
\bibitem [{\citenamefont {Fujitani}(2017)}]{preprint}%
  \BibitemOpen
  \bibfield  {author} {\bibinfo {author} {\bibfnamefont {Y.}~\bibnamefont
  {Fujitani}},\ }\href@noop {} {\bibfield  {journal} {\bibinfo  {journal} {J.
  Phys.~Soc.~Jpn.~}\ }\textbf {\bibinfo {volume} {86}},\ \bibinfo {pages}
  {044602} (\bibinfo {year} {2017})}\BibitemShut {NoStop}%
\bibitem [{\citenamefont {Fujitani}(2014{\natexlab{b}})}]{wetvisc}%
  \BibitemOpen
  \bibfield  {author} {\bibinfo {author} {\bibfnamefont {Y.}~\bibnamefont
  {Fujitani}},\ }\href@noop {} {\bibfield  {journal} {\bibinfo  {journal} {J.
  Phys.~Soc.~Jpn.~}\ }\textbf {\bibinfo {volume} {83}},\ \bibinfo {pages}
  {084401} (\bibinfo {year} {2014}{\natexlab{b}})}\BibitemShut {NoStop}%
\bibitem [{\citenamefont {Fujitani}(2016)}]{sphere}%
  \BibitemOpen
  \bibfield  {author} {\bibinfo {author} {\bibfnamefont {Y.}~\bibnamefont
  {Fujitani}},\ }\href@noop {} {\bibfield  {journal} {\bibinfo  {journal} {J.
  Phys.~Soc.~Jpn.~}\ }\textbf {\bibinfo {volume} {85}},\ \bibinfo {pages}
  {044401} (\bibinfo {year} {2016})}\BibitemShut {NoStop}%
\bibitem [{\citenamefont {Liu}\ and\ \citenamefont {Fisher}(1989)}]{liu}%
  \BibitemOpen
  \bibfield  {author} {\bibinfo {author} {\bibfnamefont {A.~J.}\ \bibnamefont
  {Liu}}\ and\ \bibinfo {author} {\bibfnamefont {M.~E.}\ \bibnamefont
  {Fisher}},\ }\href@noop {} {\bibfield  {journal} {\bibinfo  {journal}
  {Phys.~Rev.~A}\ }\textbf {\bibinfo {volume} {40}},\ \bibinfo {pages} {7202}
  (\bibinfo {year} {1989})}\BibitemShut {NoStop}%
\bibitem [{\citenamefont {Carey}\ \emph {et~al.}(1980)\citenamefont {Carey},
  \citenamefont {Scriven},\ and\ \citenamefont {Davis}}]{aiche}%
  \BibitemOpen
  \bibfield  {author} {\bibinfo {author} {\bibfnamefont {B.~S.}\ \bibnamefont
  {Carey}}, \bibinfo {author} {\bibfnamefont {L.~E.}\ \bibnamefont {Scriven}},
  \ and\ \bibinfo {author} {\bibfnamefont {H.~T.}\ \bibnamefont {Davis}},\
  }\href@noop {} {\bibfield  {journal} {\bibinfo  {journal} {AIChE J.}\
  }\textbf {\bibinfo {volume} {26}},\ \bibinfo {pages} {705} (\bibinfo {year}
  {1980})}\BibitemShut {NoStop}%
\bibitem [{\citenamefont {Cornelisse}\ \emph {et~al.}(1996)\citenamefont
  {Cornelisse}, \citenamefont {Peters},\ and\ \citenamefont
  {de~Swaan~Arons}}]{vdwexp}%
  \BibitemOpen
  \bibfield  {author} {\bibinfo {author} {\bibfnamefont {P.~M.~W.}\
  \bibnamefont {Cornelisse}}, \bibinfo {author} {\bibfnamefont {C.~J.}\
  \bibnamefont {Peters}}, \ and\ \bibinfo {author} {\bibfnamefont
  {J.}~\bibnamefont {de~Swaan~Arons}},\ }\href@noop {} {\bibfield  {journal}
  {\bibinfo  {journal} {Fluid Phase Equilib.~}\ }\textbf {\bibinfo {volume}
  {117}},\ \bibinfo {pages} {312} (\bibinfo {year} {1996})}\BibitemShut
  {NoStop}%
\bibitem [{\citenamefont {Smeulders}\ \emph {et~al.}(1990)\citenamefont
  {Smeulders}, \citenamefont {Blom},\ and\ \citenamefont {Mellema}}]{smeu}%
  \BibitemOpen
  \bibfield  {author} {\bibinfo {author} {\bibfnamefont {J.~B. A.~F.}\
  \bibnamefont {Smeulders}}, \bibinfo {author} {\bibfnamefont {C.}~\bibnamefont
  {Blom}}, \ and\ \bibinfo {author} {\bibfnamefont {J.}~\bibnamefont
  {Mellema}},\ }\href@noop {} {\bibfield  {journal} {\bibinfo  {journal}
  {Phys.~Rev.~A}\ }\textbf {\bibinfo {volume} {42}},\ \bibinfo {pages} {3483}
  (\bibinfo {year} {1990})}\BibitemShut {NoStop}%
\bibitem [{\citenamefont {Fujitani}(2011)}]{yf2011}%
  \BibitemOpen
  \bibfield  {author} {\bibinfo {author} {\bibfnamefont {Y.}~\bibnamefont
  {Fujitani}},\ }\href@noop {} {\bibfield  {journal} {\bibinfo  {journal} {J.
  Phys.~Soc.~Jpn.~}\ }\textbf {\bibinfo {volume} {80}},\ \bibinfo {pages}
  {074609} (\bibinfo {year} {2011})}\BibitemShut {NoStop}%
\bibitem [{\citenamefont {Peskin}(2002)}]{peskin}%
  \BibitemOpen
  \bibfield  {author} {\bibinfo {author} {\bibfnamefont {C.}~\bibnamefont
  {Peskin}},\ }\href@noop {} {\bibfield  {journal} {\bibinfo  {journal} {Acta
  Numer.~}\ }\textbf {\bibinfo {volume} {11}},\ \bibinfo {pages} {479}
  (\bibinfo {year} {2002})}\BibitemShut {NoStop}%
\bibitem [{\citenamefont {Machta}\ \emph {et~al.}(2012)\citenamefont {Machta},
  \citenamefont {Veatch},\ and\ \citenamefont {Sethna}}]{mach}%
  \BibitemOpen
  \bibfield  {author} {\bibinfo {author} {\bibfnamefont {B.~B.}\ \bibnamefont
  {Machta}}, \bibinfo {author} {\bibfnamefont {S.~L.}\ \bibnamefont {Veatch}},
  \ and\ \bibinfo {author} {\bibfnamefont {J.~P.}\ \bibnamefont {Sethna}},\
  }\href@noop {} {\bibfield  {journal} {\bibinfo  {journal}
  {Phys.~Rev.~Lett.~}\ }\textbf {\bibinfo {volume} {109}},\ \bibinfo {pages}
  {138101} (\bibinfo {year} {2012})}\BibitemShut {NoStop}%
\bibitem [{\citenamefont {Nowakowski}\ \emph {et~al.}(2016)\citenamefont
  {Nowakowski}, \citenamefont {Maciolek},\ and\ \citenamefont
  {Dietrich}}]{nowa}%
  \BibitemOpen
  \bibfield  {author} {\bibinfo {author} {\bibfnamefont {P.}~\bibnamefont
  {Nowakowski}}, \bibinfo {author} {\bibfnamefont {A.}~\bibnamefont
  {Maciolek}}, \ and\ \bibinfo {author} {\bibfnamefont {S.}~\bibnamefont
  {Dietrich}},\ }\href@noop {} {\bibfield  {journal} {\bibinfo  {journal} {J.
  Phys.~A: Math.~Theor.~}\ }\textbf {\bibinfo {volume} {49}},\ \bibinfo {pages}
  {485001} (\bibinfo {year} {2016})}\BibitemShut {NoStop}%
\bibitem [{err()}]{erratum}%
  \BibitemOpen
  \href@noop {} {\ }\bibinfo {note} {Accordingly, $\Lambda(\rho)$ should have
  been replaced by $\Lambda(\rho)-\chi(1)\kappa(\rho)$ in
  \citet{oldraft}}\BibitemShut {NoStop}%
\bibitem [{\citenamefont {Fujitani}(2012)}]{almost}%
  \BibitemOpen
  \bibfield  {author} {\bibinfo {author} {\bibfnamefont {Y.}~\bibnamefont
  {Fujitani}},\ }\href@noop {} {\bibfield  {journal} {\bibinfo  {journal} {J.
  Phys.~Soc.~Jpn.~}\ }\textbf {\bibinfo {volume} {81}},\ \bibinfo {pages}
  {084601} (\bibinfo {year} {2012})}\BibitemShut {NoStop}%
\bibitem [{\citenamefont {Mori}(1965)}]{mori}%
  \BibitemOpen
  \bibfield  {author} {\bibinfo {author} {\bibfnamefont {H.}~\bibnamefont
  {Mori}},\ }\href@noop {} {\bibfield  {journal} {\bibinfo  {journal}
  {Prog.~Theor.~Phys.~}\ }\textbf {\bibinfo {volume} {30}},\ \bibinfo {pages}
  {423} (\bibinfo {year} {1965})}\BibitemShut {NoStop}%
\bibitem [{\citenamefont {Kawasaki}(1970)}]{Kawa}%
  \BibitemOpen
  \bibfield  {author} {\bibinfo {author} {\bibfnamefont {K.}~\bibnamefont
  {Kawasaki}},\ }\href@noop {} {\bibfield  {journal} {\bibinfo  {journal}
  {Ann.~Phys.~(N. Y.)}\ }\textbf {\bibinfo {volume} {61}},\ \bibinfo {pages}
  {1} (\bibinfo {year} {1970})}\BibitemShut {NoStop}%
\bibitem [{\citenamefont {Zwanzig}(2001)}]{zwanzig}%
  \BibitemOpen
  \bibfield  {author} {\bibinfo {author} {\bibfnamefont {R.}~\bibnamefont
  {Zwanzig}},\ }\href@noop {} {\emph {\bibinfo {title} {Nonequilibrium
  Statistical Mechanics}}}\ (\bibinfo  {publisher} {Oxford University Press},\
  \bibinfo {year} {2001})\ \bibinfo {note} {sect.~9}\BibitemShut {NoStop}%
\bibitem [{\citenamefont {Ohta}(1975)}]{ohta}%
  \BibitemOpen
  \bibfield  {author} {\bibinfo {author} {\bibfnamefont {T.}~\bibnamefont
  {Ohta}},\ }\href@noop {} {\bibfield  {journal} {\bibinfo  {journal}
  {Prog.~Theor.~Phys.~}\ }\textbf {\bibinfo {volume} {54}},\ \bibinfo {pages}
  {1566} (\bibinfo {year} {1975})}\BibitemShut {NoStop}%
\bibitem [{\citenamefont {Ohta}\ and\ \citenamefont
  {Kawasaki}(1976)}]{ohtaKawa}%
  \BibitemOpen
  \bibfield  {author} {\bibinfo {author} {\bibfnamefont {T.}~\bibnamefont
  {Ohta}}\ and\ \bibinfo {author} {\bibfnamefont {K.}~\bibnamefont
  {Kawasaki}},\ }\href@noop {} {\bibfield  {journal} {\bibinfo  {journal}
  {Prog.~Theor.~Phys.~}\ }\textbf {\bibinfo {volume} {55}},\ \bibinfo {pages}
  {1384} (\bibinfo {year} {1976})}\BibitemShut {NoStop}%
\end{thebibliography}%

\end{document}